\documentclass[amsmath,10pt,aps,preprintnumbers,twocolumn,groupedaddress,superscriptaddress,notitlepage,nofootinbib,prd]{revtex4-1}
\usepackage[utf8]{inputenc}
\usepackage{indentfirst}
\usepackage{amssymb}
\usepackage{amsmath,latexsym}
\usepackage{graphicx}
\usepackage{float}
\usepackage[letterpaper,top=3cm,bottom=2cm,left=1.5cm,right=1.5cm]{geometry}
\usepackage{mathrsfs} % makes \mathscr into curly font
\usepackage{color}
\usepackage[dvipsnames]{xcolor}
\usepackage{soul}
\usepackage{slashed}
\usepackage{feynmf}
\usepackage{braket}
\usepackage{comment}
\usepackage{enumitem}

\usepackage{bbold}
\usepackage[bbgreekl]{mathbbol}

\usepackage{dsfont}
\newcommand{\bbid}{\mathds{1}}

\newcommand{\beq}{\begin{equation}}
\newcommand{\eeq}{\end{equation}}
\newcommand{\bea}{\begin{eqnarray}}
\newcommand{\eea}{\end{eqnarray}}

\long\def\beqs#1\eeqs{\beq\begin{split} #1 \end{split}\eeq}

\newcommand{\bo}{\mathbf}

\newcommand{\bb}{\mathbb}
\newcommand{\smallfrac}[2]{{\textstyle\frac{#1}{#2}}}

\newcommand{\BE}{\begin{equation}}
\newcommand{\EE}{\end{equation}}
\newcommand{\BA}{\begin{eqnarray}}
\newcommand{\EA}{\end{eqnarray}}
\newcommand{\mrm}{\mathrm}

\newcommand{\mi}{\mathrm{i}}
\newcommand{\me}{\mathrm{e}}
\newcommand{\mcal}{\mathcal}

\newcommand{\nn}{\nonumber}

\newcommand{\f}[1]{{\mathscr #1}}

\newcommand{\fU}{{\mathscr U}}

\newcommand{\vect}[1]{\boldsymbol{\mathbf{#1}}}

\DeclareSymbolFontAlphabet{\mathbbm}{bbold}
\DeclareSymbolFontAlphabet{\mathbb}{AMSb}

\definecolor{MyRed}{RGB}{153,0,13}

\makeatletter
\DeclareFontEncoding{LS2}{}{\noaccents@}
\makeatother
\DeclareFontSubstitution{LS2}{stix}{m}{n}
\DeclareSymbolFont{xlargesymbols}{LS2}{stixex}{m}{n}
\DeclareMathSymbol{\sumop}{\mathop}{xlargesymbols}{"B3}

 \newcommand{\sbbsquare}{
 \mbox{$
  \raisebox{0.45ex}
  {   \scalebox{0.55}{$|$} }
   \hspace{-0.64em}
  \Box
  \hspace{0.0em}
  $}
  {}}
\newcommand{\bbsquare}{\mbox{$\raisebox{-.3ex}{$\scalebox{1.3}{$\sbbsquare$}$}$}}

\usepackage[colorlinks=true,backref=none,linktocpage=true,citecolor=blue,urlcolor=blue,linkcolor=blue,pdfpagemode=UseOutlines]{hyperref}

\hypersetup{%
  bookmarksnumbered=true,
  pdftitle = {},
  pdfsubject = {},
  pdfauthor = {},
  pdfkeywords = {}
}

\usepackage{microtype}

\usepackage{pgfplotstable,array,siunitx}
\usepackage{multirow}
\usepackage{booktabs}

\usepackage{tikz}

\pgfplotsset{compat=1.18} 
\newcommand{\eq}[1]{Eq.~(\ref{#1})}
\newcommand{\fig}[1]{Fig.~\ref{#1}}

\begin{document}
\title{Fuzzy 
gauge theory for quantum computers}

\author{Andrei Alexandru}
\email{aalexan@gwu.edu}
\affiliation{Department of Physics,
The George Washington University, Washington, District of Columbia 20052}
\affiliation{Department of Physics,
University of Maryland, College Park, Maryland 20742}

\author{Paulo F. Bedaque}
\email{bedaque@umd.edu}
\affiliation{Department of Physics,
University of Maryland, College Park, Maryland 20742}

\author{Andrea Carosso}
\email{acarosso@gwu.edu}
\affiliation{Department of Physics,
The George Washington University, Washington, District of Columbia 20052}

\author{Michael J. Cervia}
\email{cervia@gwu.edu}
\affiliation{Department of Physics,
The George Washington University, Washington, District of Columbia 20052}
\affiliation{Department of Physics,
University of Maryland, College Park, Maryland 20742}

\author{Edison M. Murairi}
\email{murairi@gwu.edu}
\affiliation{Department of Physics,
The George Washington University, Washington, District of Columbia 20052}

\author{Andy Sheng}
\email{asheng@umd.edu}
\affiliation{Department of Physics,
University of Maryland, College Park, Maryland 20742}

% \preprint{}

\date{\today}

\begin{abstract}

Continuous gauge theories, because of their bosonic degrees of freedom, have an infinite-dimensional local Hilbert space. 
Encoding these degrees of freedom 
on qubit-based hardware 
demands some sort of ``qubitization'' scheme, 
where one approximates the behavior of a theory while using only finitely many degrees of freedom. 
We propose a novel 
qubitization strategy for gauge theories,  called ``fuzzy gauge theory,'' 
building on the success of the fuzzy $\sigma$-model in  earlier work. 
We provide arguments that the fuzzy gauge theory lies in the same universality class as regular gauge theory, in which case its use would obviate the need of any further limit besides the usual spatial continuum limit. 
Furthermore, we demonstrate that these models are relatively resource-efficient for quantum simulations. 

\vspace{5em}

\end{abstract}

\maketitle

\section{Introduction}
\label{sec:intro}

There are many obstacles to the use of  quantum computers  in quantum field theory  calculations. Besides the need for more reliable hardware than currently available, there are a number of conceptual questions to be answered before these calculations become feasible. This paper addresses one of these obstacles: how to reduce the infinite-dimensional Hilbert space of a bosonic field theory, in particular a gauge theory, to a finite-dimensional Hilbert space that ``fits" onto a quantum computer with finite 
registers.\footnote{The application of tensor network methods also favors theories with a finite-dimensional Hilbert space. See, for instance Ref.~\cite{Banuls:20198n}.}

Before proceeding, let us emphasize the importance of this direction and briefly 
contrast this approach to alternative numerical methods. 
The dimension of the Hilbert space in a finite-dimensional field theory grows exponentially with the volume 
($\sim D^\mathsf{V}$ for local Hilbert space dimension $D$); 
therefore, the direct diagonalization of the time evolution operator by classical computers also has a computational cost that grows exponentially with the volume $\mathsf{V}$. Monte Carlo methods are more efficient, with a cost scaling roughly as $\sim \mathsf{V}$, but they are limited to problems that can be framed as an evolution in {\it imaginary} time. This restriction leaves real-time evolution and finite-chemical potential problems, among others, mostly out of reach of Monte Carlo methods. The culprit of this limitation is the  famous sign problem that has resisted  valiant attempts to bypass it~(see Refs.~\cite{Aarts_2016,Philipsen:2007aa,deForcrand:2009zkb,10.1143/PTP.110.615,KARSCH200014,Gattringer:2016kco,Alexandru:2020wrj} for reviews in the context of QCD simulations).
The quantum simulation of the evolution can instead be performed directly in real time, manipulating a number of qubits proportional to $\mathsf{V}$ (and not exponential in $\mathsf{V}$). Therein lies a possible (exponential) quantum advantage over classical methods.

The first step of qubitization is the substitution of the continuous, $d$-dimensional physical space by a lattice with a finite lattice spacing $\mathsf{a}>0$ and finite extent $\mathsf{L}=\mathsf{a} \mathsf{N}$ in each direction. 
Space discretization is a well understood topic, since it is a routine part of the lattice field theory approach. 
Whether this discretization of spatial domain alone suffices depends on the spin statistics of one's theory. 
For purely fermionic theories, the resulting Hilbert space is already finite-dimensional, and so in principle it can readily be encoded and simulated on a quantum computer. 
However, for bosonic fields the occupation numbers at a lattice site may be arbitrarily large; even the Hilbert space of a bosonic theory defined on a single spatial site (or link) is infinite-dimensional. 
Therefore, for bosons some truncation of the {\it field} space is also required to render the {\it local} Hilbert space finite. 
This reduction of the Hilbert space to a finite-dimensional one is sometimes called ``qubitization." 
Many methods to accomplish this truncation have been proposed, which we briefly review below.

Perhaps the most obvious approach for qubitization is the substitution of the field space manifold by a finite subset: For instance, the substitution of the sphere $S^2$ in the $(1+1)$-dimensional $O(3)$ $\sigma$-model by the vertices of a platonic solid. 
The internal symmetries are reduced to a finite group, and Monte Carlo studies have shown (after substantial discussion in the literature) that these models are {\it not} in the same universality class as the $O(3)$ $\sigma$-model \cite{Hasenfratz:2001iz,Caracciolo_2001} although fairly large, but finite, correlation lengths are achievable. In the $(3+1)$-dimensional $SU(3)$ gauge theory a similar phenomenon occurs. In Ref.~\cite{PhysRevD.100.114501}, a gauge theory based on the $S(1080)$ group, the  largest (and therefore finest) ``crystal-like" discrete subgroup of $SU(3)$ was simulated. The result shows that this lattice theory does not have a continuum limit although fairly large correlation lengths can be achieved, corresponding to a lattice spacing of the order of $\mathsf{a}\approx0.08$ fm, small enough to be of phenomenological use. The reason these truncations fail to have a continuum limit can be intuited by considering the imaginary time path integral. Due to the discretization of the field values, there is a gap between the smallest and the next-to-smallest values of the action. When the coupling becomes small all field configurations except one are exponentially suppressed, resulting in the ``freezing" of the system.  
Nevertheless, there are many ongoing efforts to work around this issue in the case of $SU(2)$ gauge theory, including different samplings from the gauge group manifold~\cite{Hartung:2022hoz,Jakobs:2023lpp} as well as studies of finite subgroups~\cite{Alam:2021uuq,Gonzalez-Cuadra:2022hxt,Gustafson:2022xdt,Gustafson:2023swx,Mariani:2023eix}, $q$-deformed groups~\cite{Zache:2023dko}, and different encodings of physical degrees of freedom~\cite{Banerjee:2017tjn,Anishetty:2018vod,Raychowdhury:2018tfj,Raychowdhury:2019iki,Klco:2019evd,ARahman:2021ktn,Davoudi:2022xmb}. 

Another common approach to qubitization is to truncate the one-site (or one-link) Hilbert space to some finite irrep of the Laplacian on the target space of the theory~\cite{Barber:1980,Hamer:1982,Byrnes:2005qx,Zohar:2014qma,Liu:2021}. 
For instance, in the $O(3)$ $\sigma$-model, where the fields take value on a sphere $S^2$, this strategy 
truncates the Hilbert space 
of spherical harmonics up to some value $\ell\leq \ell_\mathrm{max}$ of the angular momentum. 
In an $SU(2)$ gauge theory, on each link one would truncate the basis 
of Wigner functions $\mathcal{D}^j_{mm'}$ up to $j\leq j_\mathrm{max}$. 
Such a truncation preserves the gauge symmetry of the theory, but it has serious drawbacks. 
First, because such a basis diagonalizes a kinetic energy operator and the truncation restricts to low irreps, one expects such a truncation to effectively capture the physics of the original theory only in a strong-coupling limit. 
In asymptotically free theories, this limit coincides with a large lattice spacing and therefore is not expected to be well-suited for studying continuum physics. 
Second, as demonstrated in the case of the $\sigma$-model, such a truncation can entirely fail to have a continuum limit outright~\cite{Alexandru:2022son}. 
Last, as a practical consideration: while such discretizations of field space can be arbitrarily improved by increasing $j_\mrm{max}$, doing so comes at the price of the number of qubits per register (i.e., site or link) and the number of gates in the simulation circuit. In fact, there is evidence that the number of gates required to simulate the time evolution grows considerably with the number of qubits per link/site~\cite{Murairi:2022zdg}. This observation suggests that there is a high cost in approaching the continuum limit that most likely negates the potential quantum advantage in such a framework.

We note that there have been alternatives to the truncation strategy described above to obtain a finite gauge theory, wherein one searches for finite matrices satisfying the exact quantum symmetry algebra~\cite{Kogut-Susskind} and constructs a gauge-invariant Hamiltonian out of those matrices. 
For example, the first such proposal was due to Horn in 1981~\cite{HORN1981149}, who explored an $SU(2)$ gauge theory with a five-dimensional one-link Hilbert space. 
The extension to higher gauge groups was considered, but then dismissed due to the presence of a spurious $U(1)$ gauge symmetry. 
Orland and Rohrlich~\cite{Orland:1989st} later proposed ``gauge magnets'' as a family of finite $SU(2)$ gauge theories, with its smallest representation being of dimension four per link. 
In 1996, the ``quantum link'' framework was introduced by Chandrasekaran and Wiese~\cite{Chandrasekharan:1996ih}, which could be formulated for arbitrary $SU(N)$~\cite{Brower:1997ha}. 
For $SU(2)$ the smallest representation was again four-dimensional, but this representation had the spurious $U(1)$ problem, as Horn's did; to remedy this complication, they proposed carrying out calculations with a six-dimensional representation for each link. 

{Regardless of the discretization strategy, one ultimately wants not only a finite-dimensional model with gauge symmetry but also a model that reproduces the spectrum of the desired continuum gauge theory in a critical limit. 
To that end, we propose a framework for ``fuzzifying'' gauge theory. 
It constitutes a generalization of the strategy we employed in the fuzzy $\sigma$-model~\cite{Alexandru:2019ozf,Alexandru:2021xkf,Alexandru:2022son}. 
Although the fuzzy approach\footnote{We stress the difference between our approach, where the {\it target} space of the theory is substituted by its fuzzy version, from the extensively studied fuzzy field theories where {\it spacetime} becomes a noncommuting space. See Ref.~\cite{Douglas_2001} for a review.}
is closer in spirit to the finite gauge theory strategy, it is guided by analogies with the original theory in the field space 
representation, entailing a different Hilbert space from the models outlined above. 
And although nonperturbative simulations of the theory are not yet accessible, we will identify throughout the paper various properties that the fuzzy model satisfies and we believe must be satisfied by any qubitization of non-Abelian gauge theories. 
For reference, these properties, together with our degrees of confidence in them \emph{vis-\`a-vis} the fuzzy model, are summarized finally in the Conclusions  (Sec.~\ref{sec:conc}). }

The bulk of this article is organized as follows. 
In Sec.~\ref{sec:review}, we briefly review the Hamiltonian formulation of untruncated gauge theories on a lattice. 
In Sec.~\ref{sec:fuzzy-algebra}, we describe the properties of ``fuzzy'' qubitizations of lattice gauge theories in general, and we consider an explicit realization in the $SU(2)$ case.  
Then, in Sec.~\ref{sec:continuum}, we explore the features of fuzzy models that allow for a continuum limit, first in the simpler case of the fuzzy $\sigma$-model, and then in the case of the fuzzy $SU(2)$ gauge theory.
Finally, in Sec.~\ref{sec:quantum-sim}, we demonstrate the simulability of the fuzzy $SU(2)$ proposal. 

\section{Lattice gauge theory in the Hamiltonian formalism}
\label{sec:review}

The fuzzy gauge theory that we propose will be formulated in the Hamiltonian language, convenient for quantum computer implementations. 
As such, we begin by reviewing some features of untruncated lattice gauge theory in this language. 
This formalism has been well-understood since Kogut and Susskind first described it~\cite{Kogut-Susskind}, and so we only briefly recapitulate it here to establish notation and highlight a few points that will be pertinent later. 
Although we will take the case of an $SU(N)$ gauge theory, the formulation readily generalizes to other compact groups. 

We start with a $d$-dimensional spatial lattice (and continuum time). 
The target space on each link is the gauge group in consideration\textemdash here an $SU(N)$ manifold (see, e.g.,~Refs.~\cite{Robson:1981,Creutz:1983} for $N=2$)\textemdash constituting the bosonic degrees of freedom of the theory. 
The wave function for a single link is a complex function of an $SU(N)$ matrix $U$, and furthermore the wave function for the total system will be a function of all links, $\psi(U_1, U_2, \ldots)$, describing both ``physical'' and ``unphysical'' states each to be distinguished  later. 
The wave function on a single link may be expanded in an infinite Taylor series
\beq\label{eq:inf-taylor}
\psi(U) = \psi_0 + \psi_{ij} U_{ij} +  \widetilde\psi_{ij} U^*_{ij} + O(U^2), 
\eeq
with implied summation over repeated color indices $i,j=1,\ldots,N$. 
Notably, these states belong to an infinite-dimensional Hilbert space, even if the spatial lattice has a finite number of links. 

A gauge transformation at a site $x$ is defined by multiplying all links connected to $x$ by an element of $ SU(N)$, from either the left or the right (depending on the orientation of the links). 
Thus, a one-link wave function $\psi$ transforms as
\begin{align}
\psi(U) & \mapsto \mathbb{L} \psi(U) = \psi(L^{-1} U), \nn\\
\psi(U) & \mapsto \mathbb{R} \psi(U) = \psi(U R),
\end{align}
depending on whether the link points away from ($L$) or into ($R$) the site $x$. 
These operators can be written in terms of infinitesimal generators, $\mathbb{T}_a^{L,R}$ ($a=1,\ldots,N^2-1$), as 
\begin{equation}
    \mathbb{L} = e^{-\mathrm{i}\, \omega^L_a  \mathbb{T}^{L}_a},
    \quad\mathbb{R} = e^{-\mathrm{i}\, \omega^R_a  \mathbb{T}^{R}_a},
\end{equation}
where real coefficients $\omega^{L,R}_a$ parametrize the group elements 
$L=\exp(- \mathrm{i} \, \omega^L_a\, \lambda_a/2),
R=\exp(-\mathrm{i}\, \omega^R_a\, \lambda_a/2)$ 
with $\frac{1}{2}\lambda_a$ being the generators of the fundamental representation of $SU(N)$. 
On a single link, the generators are Hermitian, first-order differential operators \cite{Menotti:1981,Smith:1992} satisfying
\begin{align} 
[\mathbb{T}^{L}_a, \mathbb{T}^{L}_b] & = \mathrm{i} f_{abc} \mathbb{T}^{L}_c, 
\label{eq:l-gen-comm} \\
[\mathbb{T}^{R}_a, \mathbb{T}^{R}_b] & = \mathrm{i} f_{abc} \mathbb{T}^{R}_c, 
\label{eq:r-gen-comm} \\
[\mathbb{T}^{L}_a, \mathbb{T}^{R}_b] & = 0, 
\label{eq:lr-gen-comm} 
\end{align}
where $f_{abc}$ is the totally antisymmetric structure tensor of $SU(N)$. 

We define field operators $\mathbb{U}_{ij}$ at every link by
\beq
\mathbb{U}_{ij} \psi(U) = U_{ij} \psi(U),
\eeq
which implies the commutation relations
\beq\label{eq:U-commute}
[\mathbb{U}_{ij}, \mathbb{U}_{kl}] = [\mathbb{U}_{ij}, \mathbb{U}^*_{kl}] =0.
\eeq 
Now, the  links $U$ are $SU(N)$ matrices whose components obey the relations
\begin{align}
U^*_{ji}U_{jk} = \; U_{ij} & U^*_{kj} = \delta_{ik}, \label{eq:unitarity}
\\ 
\det U = \; & 1. \label{eq:special}
\end{align} 
The same relations are therefore satisfied by the field operators $\mathbb{U}_{ij}, \mathbb{U}^*_{ij}$, and the ordering of the operators is immaterial since $\mathbb{U}_{ij}, \mathbb{U}^*_{ij}$ commute among themselves.

From the previous definition of gauge transformations, one finds that the link operators transform as 
\BE
\mathbb{R}^\dagger \, \mathbb{L}^\dag  \, \mathbb{U}_{ij} \, \mathbb{L} \, \mathbb{R} = L_{ii'} \mathbb{U}_{i'j'} (R^{-1})_{j'j},
\label{eq:ord-u-transf}
\EE
the infinitesimal versions of which satisfy 
\begin{align} %\label{eq:U-sym-comms}
[\mathbb{T}^{L}_a, \mathbb{U}_{ij}] &=  -\frac{1}{2}\left(\lambda_a\right)_{ii'} \mathbb{U}_{i'j}, 
\label{eq:lU-comm} \\
[\mathbb{T}^{R}_a, \mathbb{U}_{ij}] &=  +\frac{1}{2}\mathbb{U}_{i j'}\left(\lambda_a\right)_{j'j}.
\label{eq:rU-comm}
\end{align}
Finally, the physical Hilbert space of the theory is defined as the gauge-invariant subspace of the total Hilbert space, that is, the wave functions satisfying
\begin{align}
e^{-\mi \omega_a\mathbb{G}_a(x)} \psi_\mathrm{phys}(\{U_\ell\}) = \psi_\mathrm{phys}(\{U_\ell\}), \label{eq:nonfuzz-gauss-constraint} \\
\mathbb{G}_a(x)
=
\sum_\mu \big[\mathbb{T}^{L}_a(x,\mu) + \mathbb{T}^{R}_a(x-\hat{\mu},\mu) \big] 
\label{eq:nonfuzz-gauss-op}
\end{align} 
for all $x$ and $a$; here, $\mu$ runs over all spatial directions. The infinitesimal condition $\mathbb{G}_a(x) \psi_\mathrm{phys}(\{ U_l \})=0$ is the quantum version of Gauss's law.

Using Eqs.~\eqref{eq:l-gen-comm} and \eqref{eq:r-gen-comm} as well as \eqref{eq:lU-comm} and \eqref{eq:rU-comm}, we may confirm the gauge invariance of the Kogut-Susskind Hamiltonian,
\begin{align}
\mathsf{a}\mathbb{H} &= \frac{g^2}{2} \sum_{\ell} \big[ \mathbb{T}^{L}(\ell)^2 +  \mathbb{T}^{R}(\ell)^2 \big]
 -\frac{1}{2g^2} \sum_{P} \bbsquare(P) + \mrm{H.c.},\label{eq:H-KS}
\end{align} 
where $\ell=(x,\mu)$ label the links of the lattice and $P=(x,\mu,\nu)$ label the plaquettes,
\BE
\bbsquare(P) = \mathbb{U}_{ij}(\ell_1) \, \mathbb{U}_{jk}(\ell_2) \, \mathbb{U}_{lk}(\ell_3)^* \, \mathbb{U}_{il}(\ell_4)^* . 
\EE 
Here, $\mathbb{U}(\ell_i)$ are the positively oriented links comprising a plaquette $P$. 
We work in units of the lattice spacing, so that $g^2$ is dimensionless for any chosen spacetime. For $d < 3$, the dimensionful bare coupling is $\mathsf{a}^{d-3}g^2$, but the continuum limit is achieved by $g^2 \to 0$. 
Of course, many other gauge-invariant terms may be added to Eq.~\eqref{eq:H-KS}, but such modified Hamiltonians are believed to lead to the same continuum limit as $g^2\rightarrow 0$.\footnote{To recover Lorentz invariance in the continuum limit, one should multiply the operators in $\mathbb{H}$ by an appropriately determined renormalization constant $\eta(g^2)$ \cite{Shigemitsu:1981}, as also described in 
Ref.~\cite{Alexandru:2021xkf}.} Taking this limit is a nontrivial process, which has been studied extensively. 
To summarize that analysis, it is found that as $g^2\rightarrow 0$, correlation lengths $\xi$ become exponentially larger than the lattice spacing [$\xi/\mathsf{a} \sim  \exp(1/g^2)$] and the continuum limit is thus achieved.  
Additional terms in the Hamiltonian, {\it as long as they do not break any symmetry}, only change the rate at which this same limit is approached, on account of the universality of critical phenomena \cite{RevModPhys.47.773}. 
These additional terms are sometimes used to achieve higher accuracy with coarser lattices (i.e., the so-called Symanzik improvement program \cite{Symanzik:1983,PARISI198558}).

The ingredients we have presented, namely a Hilbert space, a realization of gauge transformations on this space, and consequently the structure of typical gauge-invariant observables such as the Hamiltonian, are enough to define a quantum gauge theory. 
Note that the commutativity of field operators, as expressed in \eq{eq:U-commute}, did not play an essential role in the (gauge) symmetry properties of the theory.

\section{Fuzzy lattice gauge theory}
\label{sec:fuzzy-algebra}

Our goal is to define a theory with a finite-dimensional Hilbert space having the same continuum limit as that of the $SU(N)$ gauge theory defined in the previous section. 
To this end, we borrow a strategy from the philosophy of noncommutative geometry: at each link of the lattice, we replace the Hilbert space of functions on $SU(N)$ by a \textit{finite-dimensional space of matrices}, in which each matrix element $U_{ij}$ is replaced by a $D\times D$ matrix $\f U_{ij}$ for each $i,j=1,\ldots,N$, with $D$ yet to be determined.
If the $\f U_{ij}$, together with their conjugate-transposes $(\f U_{ij})^\dag$ and products thereof, span the space of matrices, then states on this ``fuzzy $SU(N)$'' will be expressible as a series, analogous to \eq{eq:inf-taylor},
\beq \label{eq:fuzzy-taylor}
\Psi = \Psi_0 \openone + \Psi_{ij} \f U_{ij} + \widetilde\Psi_{ij} (\f U_{ij})^\dagger + O(\fU^2),
\eeq
where the complex numbers $\widetilde \Psi_{ij}$ are again not necessarily related to $\Psi_{ij}$. 
Because sufficiently high powers of these finite matrices will not be linearly independent from lower powers, the series must terminate at a finite order. 
The Hilbert space on each link has the inner product $\langle \Psi | \Phi \rangle := \mrm{tr}[ \Psi^\dag \Phi ]
$ and  dimension $D^2$.

At a minimum, the finite theory should have the same internal symmetries as the untruncated theory, according to the standard universality arguments of the Wilsonian renormalization group. 
In particular, 
we must find $D\times D$ matrices $\mathcal{L}$ and $\mathcal{R}$ that realize gauge transformations via conjugation:
\beq
\Psi \; \mapsto \; \mathcal{R} \, \mathcal{L} \, \Psi \, \mathcal{L}^\dagger \, \mathcal{R}^\dagger.
\label{eq:fuzzy-state-transformed}
\eeq
The generators of these transformations are given by 
\begin{equation} 
\mathcal{T}^{L}_a = [T^{L}_a, \bullet],
\quad\mathcal{T}^R_a = [T^{R}_a, \bullet],
\label{eq:fuzzy-transform-gen}
\end{equation}
where ${T}^{L,R}_a$ are $D$-dimensional representations of the $\smallfrac{1}{2} \lambda_a$, satisfying the algebra  
Eqs.~(\ref{eq:l-gen-comm}-\ref{eq:lr-gen-comm}). From this one can readily check that $\mathcal{T}^{L,R}_a$ also satisfy this same algebra. 
To constitute a gauge symmetry, the operators $\mathcal{T}^{L,R}_a$ and $\f U_{ij}$ together must also satisfy analogs of Eqs.~\eqref{eq:lU-comm} and \eqref{eq:rU-comm}. 
This property will be guaranteed as long as ${T}^{L,R}_a$ themselves with $\f U_{ij}$ obey those equations, 
owing to the Leibniz rule obeyed by commutators $\mathcal{T}^{L,R}_a$. Thus, for any matrices $\fU_{ij}, \; T_a^{L,R}$ satisfying a $D$-dimensional representation of the gauge symmetry algebra, the operators $\fU_{ij}, \mcal T_a^{L,R}$ will satisfy an (adjoint) representation of that same algebra,
\begin{align}
    [\mathcal{T}^L_a,\mathcal{T}^L_b] &= \mathrm{i} f_{abc} \mathcal{T}^L_c, \\
    [\mathcal{T}^R_a,\mathcal{T}^R_b] &= \mathrm{i} f_{abc} \mathcal{T}^R_c, \\
    [\mathcal{T}^L_a,\mathcal{T}^R_b] &= 0, \\
    [\mathcal{T}^L_a,\fU_{ij}] &= -\frac{1}{2}\left( \lambda_a\right)_{ii'}\fU_{i'j}, \\
    [\mathcal{T}^R_a,\fU_{ij}] &= +\frac{1}{2}\fU_{ij'}\left(\lambda_a\right)_{j'j},
\end{align}
when acting on the fuzzy Hilbert space. It follows, for example, that the link matrices transform as
\begin{align} \label{g-trans}
e^{-\mathrm{i}\,\omega^R_a \mathcal{T}^{R}_a} e^{-\mathrm{i}\,\omega^L_a \mathcal{T}^{L}_a} \fU_{ij} &= \mathcal{R} \, \mathcal{L} \, \fU_{ij} \, \mathcal{L}^\dagger \, \mathcal{R}^\dagger \nonumber \\
&= L_{ii'}\fU_{i'j'}(R^{-1})_{j'j},
\end{align} 
with $\mathcal{L}=\exp(-\mathrm{i}\,\omega^L_a {T}^{L}_a)$ and $\mathcal{R}$ defined analogously. From these matrices, one may construct gauge-invariant operators in the same way as in the untruncated gauge theory (i.e., by contracting color and adjoint indices appropriately). 

For lattice gauge theories or chiral models with target space $SU(N)$, the ``classical'' elements $U_{ij} \in \bb C$ satisfy special unitarity Eqs.~\eqref{eq:unitarity} and \eqref{eq:special} defining the $SU(N)$ manifold. 
In a fuzzy theory, one seeks matrices $\f U_{ij}$ which satisfy analogs of these equations, though it is clear that the noncommutativity of matrices implies that there may not be a unique way to define the fuzzy manifold counterpart to Eqs.~\eqref{eq:unitarity} and \eqref{eq:special}. 
For example, a natural choice would be to find matrices satisfying the set of $N^2$ equations
\begin{align} \label{typeI}
\frac{1}{2} \big\{ (\fU_{ji})^\dagger, \fU_{jk} \big\} &= \delta_{ik} \bbid \nn\\
\mathrm{or}\quad \frac{1}{2} \big\{ \fU_{ij}, (\fU_{kj})^\dag \big\} &= \delta_{ik} \bbid,
\end{align}
where $\{\bullet,\bullet\}$ is the anticommutator, and we have regarded $(\fU_{ij})^\dag$ as the noncommutative counterpart to $U_{ij}^*$.
By using the anticommutator, we have totally symmetrized the products that occur in the classical manifold definition. 
We also stress that the unitarity constraints themselves are gauge-invariant, as one observes by applying the transformation Eq.~\eqref{g-trans}.
One can furthermore seek operator-valued manifold relations corresponding to the classical determinant condition (up to a chosen symmetrization of the products of $\fU$s), 
\begin{equation}
\frac{1}{N!} \varepsilon_{i_1\cdots i_N}\varepsilon_{j_1\cdots j_N} \fU_{i_1j_1}\cdots \fU_{i_Nj_N}
= \bbid,\label{eq:fuzzy-special}
\end{equation}
where $\varepsilon$ is the Levi-Civita pseudotensor of rank $N$. 
(In Appendix~\ref{sec:fun-in-the-un}, we describe a family of fuzzy models satisfying a different kind of fuzzy unitarity.)

For the case of $N=2$, we can parametrize 
\begin{align}
    \f U_{ij} = \delta_{ij} \Gamma_0 - \mathrm{i} (\sigma_a)_{ij} \Gamma_a, 
    \label{eq:fU-gamma}
\end{align}
with Hermitian matrices $\Gamma_0$ and $\Gamma_a$, $a=1,2,3$, analogous to the real coefficients of $U_{ij} = \delta_{ij} x_0 - \mi (\sigma_a)_{ij} x_a$. 
A series of representations of $\f U_{ij}, T_a^{L,R}$ (or, alternatively, $\Gamma_0, \Gamma_a,T_a^{L,R} $)  were worked out in Ref.~\cite{Orland:1989st}. There it was shown that the ten operators $T^L_a$, $T^R_a$, $\Gamma_0$, and $\Gamma_a$ can be recognized as the ten generators of $\mathrm{SO}(5)$ and thus every irrep of $\mathrm{SO}(5)$ defines a representation of Eqs.~\eqref{eq:l-gen-comm}-\eqref{eq:lr-gen-comm}, \eqref{eq:lU-comm}, and \eqref{eq:rU-comm}. Just as well, $\mcal T^{L,R}_a, \f U_{ij}$ are also then specified by these irreps, by definition, as per Eqs.~\eqref{eq:fuzzy-transform-gen} and \eqref{eq:fU-gamma}. For instance, 
the four-dimensional (i.e., spinorial) representation of $\mathrm{SO}(5)$, leads to 
\bea\label{eq:d4}
\Gamma_0 &=& \frac{1}{2}\begin{pmatrix} 
0 & \openone\\ 
\openone &  0 \end{pmatrix}, \quad
\Gamma_a = \frac{1}{2} 
\begin{pmatrix} 
0 & \mathrm{i}\sigma_a\\-\mathrm{i}\sigma_a &  0 \end{pmatrix}, \nn\\  
T_a^R &=& \frac{1}{2} 
\begin{pmatrix}
\sigma_a & 0 \\ 
0 & 0  \end{pmatrix}, \qquad
T_a^L = \frac{1}{2} 
\begin{pmatrix} 
0 &  0 \\ 
0 &  \sigma_a \end{pmatrix}.
\eea 
In this representation, fuzzy analogs of the special unitarity constraints, Eqs.~\eqref{typeI} and \eqref{eq:fuzzy-special}, 
are satisfied. 
The wave function (of a single  link) can then be expanded, for example, as
\begin{align}
\Psi = &\Psi_0 \openone + \Psi_5 \Gamma_5 + \Psi_\alpha \Gamma_\alpha + \mathrm{i}\,\Psi_{5\alpha} [\Gamma_5,\Gamma_\alpha] + \mathrm{i}\,\Psi_{\alpha\beta} [\Gamma_\alpha, \Gamma_\beta],
\end{align}
where $\Gamma_5 = 8\,\Gamma_1\Gamma_2\Gamma_3\Gamma_0$ and therefore has the form of Eq.~\eqref{eq:fuzzy-taylor}. 
This example forms a 16-dimensional Hilbert space (per link), in contrast with the original, four-dimensional (i.e., vector-valued) representation of the model in Ref.~\cite{Orland:1989st}. 
Higher powers of $\Gamma_\alpha$ are not linearly independent from the lower ones, so the series terminates as promised. 

We emphasize that, contrary to usual gauge theories, the field operators $\f U_{ij}$ and $ (\f U_{ij})^\dagger$ here do not commute among themselves, and our Hilbert space is itself comprised by finite-dimensional matrices. 
As remarked above, this property has no bearing on the gauge symmetry of the theory. Heuristically, it is as if $\Psi$ could be viewed as a ``function" of noncommuting variables $\f U_{ij}$ in a way analogous to how $\psi(U)$ is a function of the commuting variable $U_{ij}$.

The construction thus far guarantees the usual gauge symmetry but does not exclude the possibility of the theory having a larger symmetry. In particular, past (finite) Hamiltonian formulations of gauge theories, such as Refs.~\cite{HORN1981149,Brower:1997ha}, contain an extra $U(1)$ symmetry.
In ordinary $SU(N)$ lattice gauge theory, the links $U$ cannot be transformed under $U(1)$ without carrying them out of the target space $SU(N)$. 
However, in a finite gauge theory, the finite-dimensional link operators are not $SU(N)$-valued, so it is possible that there exist matrices that generate left and right $U(1)$ transformations. 
This property would make the theory possess an unintended $U(1)$ gauge symmetry. 
Such a situation leads one to either \textit{break} the extra $U(1)$ explicitly in the Hamiltonian or otherwise devise a method to project to the trivial irrep of $U(1)$ on each link. 
Nevertheless, one can prove that in our case there do not exist matrices that generate a $U(1)$ gauge symmetry, in the sense that there are no matrices $T^{R,L}_0$ satisfying
$$e^{-\mathrm{i}\omega T^{R,L}_0} \, \fU_{ij} \, e^{\mathrm{i}\omega T^{R,L}_0} = e^{\mp\mathrm{i}\omega}\fU_{ij},$$
while commuting with the $SU(2)$ generators.

Just as in the regular gauge theory, the gauge-invariant physical states form a subspace of the total Hilbert space defined by
\begin{align}
e^{-\mi \omega_a \mathcal{G}_a(x)} & \Psi_{phys} =  \Psi_{phys},
\label{eq:fuzzy-phys}
\end{align}
where
\begin{align}
\mathcal{G}_a(x) = \sum_{\mu=1}^d \big( \mathcal{ T}^{L}_a(x,\mu) + \mathcal{T}^{R}_a(x-\hat{\mu},\mu) \big) 
     = [G_a(x),\bullet] \label{eq:fuzzy-gauss}
\end{align}
with $G_a(x)$ defined as in the ordinary theory, Eqs.~\eqref{eq:nonfuzz-gauss-constraint} and \eqref{eq:nonfuzz-gauss-op}, with $\mathbb T_a^{L,R} \to T_a^{L,R}$. 
We remark that in a fuzzy gauge theory, the state with identity matrices on all links is always a trivially gauge-invariant state. A family of gauge-invariant states can be constructed from this trivial singlet by applying invariant products of $\fU$s. In fact, any invariant state of the \textit{untruncated} theory can be directly mapped into a fuzzy state, and therefore previous enumerations of the gauge-invariant states of the full theory may be of use \cite{Robson:1981, Burgio:1999}. 
However, only a finite subset of such states will be linearly independent, due to the closure of the matrix algebra. 
Furthermore, the fuzzy gauge theory may have \textit{more} gauge-invariant products of field operators than those enumerated by the family above (up to closure redundancy), due to the noncommutativity of the $\fU_{ij}$s.

The construction of possible Hamiltonian operators for a fuzzy gauge theory proceeds by analogy with the Kogut-Susskind Hamiltonian. The immediate choice is to take
\BE
\mathsf{a}\mathcal{H} = \frac{\eta g^2}{2} \sum_{\ell} \mathcal{K}(\ell) \pm \frac{\eta}{2g^2} \sum_{P} \square(P)+\mathrm{H.c.}, 
\label{eq:ham-generic}
\EE
where $\square$ is the plaquette term
\begin{align}
\square(P) = \fU_{ij}(\ell_1) \fU_{jk}(\ell_2) \fU_{lk}(\ell_3)^\dag \fU_{il}(\ell_4)^\dag,
\label{eq:fuzzy-plaquettes}
\end{align}
and $\mathcal{K}$ the kinetic term
\bea \label{kinetic-1}
\mathcal{K}_1 = (\mathcal{T}^{L})^2 + (\mathcal{T}^{R})^2 = 
[T^L_a, [T^L_a, \bullet]] + [T^R_a, [T^R_a ,\bullet]]
\eea 
on each link. Above, $\ell_i$ indexes the links of an elementary plaquette $P$ on a lattice, and each link operator $\fU_{ij}$ acts via left multiplication, for reasons we explain in the next section. 
{The coupling $g^2$ is taken to be dimensionless. If the model has 
a critical point as $g^2 \to 0$, then we expect $g^2$ to be closely related to the bare coupling of the original theory.}
The multiplicative factor $\eta(g^2)$ is included to allow for the recovery of Lorentz covariance in the event that the above Hamiltonian has a relativistic dispersion relation.
Other gauge-invariant terms can be added to the Hamiltonian. For instance, 
\begin{align} 
\mathcal{K}_2 & = - \fU_{ij} \bullet (\fU_{ij})^\dag - (\fU_{ij})^\dag \bullet \fU_{ij}, \label{eq:kin-u} 
\end{align}
\begin{align}
\mathcal{K}_3 & = \Gamma_5 \label{eq:fuzz-gamma5}
\end{align} 
can be used as kinetic terms instead of (or in addition to) $\mathcal{K}_1$.\footnote{
We remark that fuzzy unitarity, such as in Eq.~\eqref{typeI}, combined with the demand of time-reversal invariance, implies that the seemingly viable kinetic term $\mcal K_4=\fU_{ij} (\fU_{ij})^\dag + (\fU_{ij})^\dag \fU_{ij} \propto \bbid$ does not alter the dynamics of the theory. 
} 
Similarly, one can add terms involving the product of $\f U_{ij}$ over Wilson loops larger than a single plaquette. 
Nevertheless, the only thing that is required of the terms included is that, {\it in the continuum limit}, they reproduce an $SU(N)$ gauge theory, and experience with regular lattice gauge theories strongly suggests that the addition of larger Wilson loops to the Hamiltonian does not change the universality class of the theory. 
Still, the question of whether the fuzzy theory (with one or more of the kinetic terms above) reduces to regular gauge theory in the continuum limit is a delicate question, to be discussed in the next section.

Now that we have defined a Hamiltonian model as well as our Hilbert space, we may also observe that our fuzzy $SU(2)$ gauge theory carries a certain \emph{global} center $\mathbb{Z}_2$ symmetry \cite{PhysRevD.24.450}.\footnote{We thank Aleksey Cherman for asking insightful questions on this point.} 
Let us define a center transformation by $\mathcal{Z}_S=\prod_{\ell\perp S}\mathcal{Z}(\ell)$, where 
\begin{align} 
\mathcal{Z}(\ell) = e^{-\mathrm{i}2\pi\mathcal{T}^L_a(\ell)} = \gamma_5(\ell)\bullet\gamma_5(\ell),
\end{align}
acting via conjugation by $\gamma_5$ only on the links intersecting with an affine [i.e., $(d-1)$-dimensional] hyperplane of the spatial lattice. 
Two important identities of this operator are $\mathcal{Z}_S\fU_{ij}(\ell)\mathcal{Z}_S^\dagger = s_\ell\fU_{ij}(\ell)$, with $s_\ell=-1$ if $\ell\perp S$ and $s_\ell=+1$ otherwise, and $\mathcal{Z}_S$ commutes with all $\mathcal{T}^{L/R}_a$. 
From the second identity, we  conclude that the global center symmetry transformation commutes with the local fuzzy gauge transformations generated by the operators of Eq.~\eqref{eq:fuzzy-gauss}. 
From the first identity, we can see that an elementary square plaquette is invariant under such transformations, while a Wilson loop intersecting with this hyperplane on a lattice with periodic boundary conditions will flip sign. 
Meanwhile, the possible kinetic terms $\mathcal{K}_1$, $\mathcal{K}_2$, and $\mathcal{K}_3$ are all invariant under the center transformation. 
In summary, the Hamiltonian of our fuzzy $SU(2)$ gauge theory possesses a global center symmetry just as in the untruncated theory. 

% Greensite pg. 12-13, Ref 11 = Polyakov: Gauge Fields and Strings
% Greensite claims: Unbroken Z2 symmetry = confinement, on a lattice with finite time extent.

\section{The continuum limit of fuzzy theories}
\label{sec:continuum}

Whether the fuzzy gauge theory defined above is equivalent to ordinary gauge theory in the continuum limit is a dynamical question that, just as most dynamical questions in non-Abelian gauge theories, is difficult to settle. 
In order to gain some insights into this question, let us first consider the fuzzy qubitization of the better-understood $(1+1)$-dimensional $O(3)$ nonlinear $\sigma$-model, and identify the salient features indicating the correct universality class. 
We then discuss the corresponding features in the fuzzy gauge theory.

\subsection{\texorpdfstring{$\sigma$}{Lg}-model}
\label{sec:continuum-sigma}

The lattice $O(3)$ $\sigma$-model can be defined as follows. 
Consider a one-dimensional spatial lattice with $\mathsf{N}=\mathsf{L}/\mathsf{a}$ sites. 
The local Hilbert space defined at each site $x$ is the set of square-integrable complex functions of a three-dimensional unit vector $ \vect{n}$: $\psi(\vect{n})$. 
These states can be expanded in an infinite series of the components $\vect{n}_i$ ($i=1,2,3$):
\beq\label{eq:sigma-taylor}
\psi(\vect{n}) = \psi_0 + \psi_i \vect{n}_i + O(\bo n_i \bo n_j).
\eeq
The global Hilbert space of the theory is the tensor product of all one-site Hilbert spaces.
We define ``position operators" $\mathbbm{n}(x)$ at every site $x$ as multiplication by $\bo n$: $\mathbbm{n}(x) \psi(\{ \vect{n}(x)\}) = \vect{n}(x)\psi(\{ \vect{n}(x) \}) $. 
A (global) $O(3)$ rotation $R$ is implemented on the wave function by the operator $\mathbb{R}$ defined by $\mathbb{R} \psi(\{ \vect{n}(x) \}) = \psi(\{R^{-1} \vect{n}(x) \})$, which can be written in terms of generating operators as $\mathbb{R}=\exp(-\mi\omega_k\mathbb{T}_k)$.
In the ``position space'' representation, $\bb T_k$ are differential operators on the sphere. The operators $\mathbbm{n}$ and $\mathbb{T}$ at one site satisfy the algebra:
\begin{align}
    \label{eq:O3comm-nn}
[ \mathbbm{n}_i, \mathbbm{n}_j] &= 
0,  \\
[ \mathbb{T}_i, \mathbb{T}_j ] &= \mathrm{i} \epsilon_{ijk} \mathbb{T}_k, \label{eq:O3comm-TT}
 \\ 
[ \mathbb{T}_i, \mathbbm{n}_j ] &= \mathrm{i} \epsilon_{ijk} \mathbbm{n}_k,
\label{eq:O3comm-Tn}
\end{align} 
while operators at distinct sites commute. 
Additionally, since $\bo n$ are unit vectors, the operators obey the constraints 
\begin{equation}
   \big( \mathbbm{n}(x) \cdot \mathbbm{n}(x) -\openone\big) \psi = 0,
    \label{eq:sigma-constraint}
\end{equation}
for all states $\psi$ and every site $x$.
A Hamiltonian invariant under $O(3)$ is

\beq
\mathsf{a}\mathbb{H} = \eta g^2 \sum_{x} \mathbb{T}(x)^2 - \frac{\eta}{g^2} \sum_x \mathbbm{n}(x) \cdot \mathbbm{n}({x+1}).
\eeq 
The continuum limit is obtained as $g^2\rightarrow 0$, where the correlation lengths diverge (in units of the lattice spacing). 
In order to recover Lorentz invariance, the parameter $\eta$, which alters the energy levels but not the wave function or correlation length $\xi$, has to be tuned so that the energy gap equals $1/\xi$. 
The continuum theory is solvable \cite{ZAMOLODCHIKOV1978525} and describes a triplet of particles with mass $M=1/\xi$ interacting elastically with the phase shift of a $\delta$-distribution potential. 
There are two limiting  energy scale regimes to consider: $E \ll M\ll 1/\mathsf{a}$ (infrared) and $M \ll E \ll 1/a$ (ultraviolet).\footnote{The $E\agt 1/\mathsf{a}$ regime is not universal and consequently is of no interest.}  
In the ultraviolet regime the theory is asymptotically free, similar non-Abelian gauge theories. 

As mentioned in Sec.~\ref{sec:intro}, one can try to qubitize this theory by substituting the target space (the unit sphere $S^2$) by a finite set of points equally spaced over the sphere (i.e., the vertices of Platonic solids). 
This approach lacks a continuum limit \cite{Caracciolo_2001,Patrascioiu_1998,Hasenfratz:2001iz}, although large finite correlation lengths can be found. 
An analogous approach is taken in Refs.~\cite{Alexandru:2019nsa,Alexandru:2021jpm} for $SU(3)$ gauge theories and leads to similar results. 
Another approach would be to restrict the local Hilbert space at every site to functions that can be expanded in terms of spherical harmonics up to some value $\ell \leq \ell_\mathrm{max}$: $\psi(\vect{n}) = \sum_{\ell=0}^{\ell_\mathrm{max}} \psi_{\ell m} Y_{\ell m}(\vect{n})$. 
In Ref.~\cite{PhysRevD.99.074501} evidence was found that the $\sigma$-model is recovered in the limit $\ell_\mathrm{max}\rightarrow \infty$, followed by the $g^2\rightarrow 0$ limit.
% Inconveniently for practical calculations, 
Taking the continuum limit requires taking larger $\ell_\mathrm{max}$ truncations, since it is known from
Ref.~\cite{Alexandru:2022son} that at fixed $\ell_\mathrm{max}$ the theory does not have a critical point. 

The final $\sigma$-model regularization that we consider is the fuzzy qubitization, whose gauge theory generalization we have outlined in previous sections. 
This model was proposed in Refs.~\cite{Alexandru:2019ozf,Singh:2019uwdsuu} and further studied in Refs.~\cite{Alexandru:2021xkf,Alexandru:2022son,Bhattacharya:2020gpm,Liu:2021}. 
It exhibits a high degree of universality which suggests that the continuum $\sigma$-model can be recovered in an appropriate limit, even at finite regularization. 

Here, we substitute the local Hilbert space of the $\sigma$-model with a Hilbert space of dimension $(2j+1)^2$, where $j$ indexes irreps of $SU(2)$: $j=0,\frac{1}{2},1,\ldots$. The elements of every local Hilbert space are $(2j+1)\times (2j+1)$ matrices. The (one-site) wave functions can be expanded as
\beq
\Psi = \psi_0 \openone+ \psi_k T_k + O(T^2) ,
\eeq 
where $T_k$ are the generators of the $j$ irrep of $SU(2)$.
In contrast with the expansion in \eq{eq:sigma-taylor}, this series terminates. 
$O(3)$ rotations act as 
\begin{equation}
\Psi \mapsto \mathcal{R}^\dagger \Psi \mathcal{R} =
e^{-\mi \omega_k\mathcal{T}_k } \Psi,
\end{equation} 
where 
$\mathcal{T}_k = [T_k, \bullet]$. 
Since the $T_k$ satisfy $[T_i, T_j] = \mi \epsilon_{ijk}T_k$, so do $\mathcal{T}_k$. 
The analogs to the position operators are
\beq
\mathfrak{n}_k = \frac{1}{\sqrt{j(j+1)}} T_k,
\eeq
which act via left multiplication. 
Because $\mathfrak{n}^2 = T^2/j(j+1) = \openone$, all states in the Hilbert space satisfy \eq{eq:sigma-constraint}. 
Now, whereas $[ \mathbbm{n}_i, \mathbbm{n}_j] = 0$, the $\mathfrak{n}_i$ do not commute:
\begin{equation}
[\mathfrak{n}_i, \mathfrak{n}_j] = \frac{1}{\sqrt{j(j+1)}} \mi \epsilon_{ijk} \mathfrak{n}_k.
\end{equation} 
The Hamiltonian is
\begin{align}\label{eq:fuzzysphere}
\mathsf{a}\mathcal{H}
   &={\eta g^2} \sum_x \mathcal{T}(x)^2 \pm \frac{\eta}{g^2} \sum_x \mathfrak{n}(x)\cdot \mathfrak{n}(x+1),
\end{align} 
where the neighbor operator is chosen to act on states from the left.
The construction we have given is just the well-known construction of the fuzzy sphere \cite{hoppe2002membranes,Madore:1991bw}. 
In the limit $j\rightarrow\infty$, the operator algebra of the fuzzy $\sigma$-model reduces to that of the untruncated lattice $\sigma$-model, i.e., Eqs.~\eqref{eq:O3comm-nn}-\eqref{eq:O3comm-Tn}. 

A more relevant consideration is whether, at \textit{fixed} $j$, the fuzzy model has a continuum limit that matches the continuum $\sigma$-model. We focus on $j=1/2$ in the following discussion. 
To study this problem let us use the basis of matrices 
$\mathrm{E}_{ab} = \mathrm e_a \otimes\mathrm e_b$, with
\beq
\mathrm e_1 =
\begin{pmatrix}
1\\
0%\\
% 0\\
% \vdots\\
% 0\\
\end{pmatrix},
\qquad
\mathrm e_2 =
\begin{pmatrix}
0\\
1
\end{pmatrix},
\eeq
and expand the (one-site) wave function in this basis: $\Psi = \psi_{ab} \mathrm{E}_{ab}$. 
The result of acting on a state with $\mathfrak{n}_k$ is 
\beq
\mathfrak{n}_k \Psi = 
\frac{2}{\sqrt{3}}
(T_k)_{aa'}\psi_{a'b} \mathrm{E}_{ab},
\eeq while the $\mathcal{T}_k$ operator involves multiplication from the left and from the right
\beq
\mathcal{T}_k \Psi = [T_k, \Psi] = 
\big((T_k)_{aa'}\psi_{a'b}-\psi_{ab'}(T_k)_{b'b} \big)\mathrm{E}_{ab}.
\eeq 
Meanwhile, since the neighbor operator $\mcal V = \sum_x \mathfrak{n}(x)\cdot \mathfrak{n}(x+1)$ involves only left multiplication, its eigenstates factorize as
\begin{eqnarray} \label{V-eigstates}
\Psi = (\psi_n)_{a_1 \cdots a_\mathsf{N}}  
\chi_{b_1 \cdots b_\mathsf{N}}  \mathrm{E}_{a_1b_1} \otimes \cdots  \otimes \mathrm{E}_{a_\mathsf{N}b_\mathsf{N}}
\equiv \ket{n} \otimes \ket{\chi},
\end{eqnarray} 
where $\ket{n}$ is an energy eigenstate of the spin-$1/2$ Heisenberg chain, while $\chi_{b_1 \cdots b_\mathsf{N}} $ can be any $2^\mathsf{N}$-dimensional state. 
(Notably, this feature is not guaranteed if the neighbor operator acts by a mixture of left and right multiplication.) Thus, there is a $\geq2^\mathsf{N}$-degenerate family of eigenstates of $\mcal V$ for each $n$.

The factorization property of $\mcal V$ can also be understood as follows: each site $x$ sees an ``$a$'' (``$b$'') spin acted on via left- (right-)multiplication operations. 
Therefore, in this basis, the one-dimensional $j=1/2$ fuzzy chain is seen to be equivalent to the Heisenberg comb of Ref.~\cite{Singh:2019uwdsuu}, which consists of two spins-$1/2$ at every site. 
Specifically, the neighbor operators in $\mathcal{V}$ couple neighboring $a$ spins, while the kinetic operator $\mathcal{T}(x)\cdot\mathcal{T}(x)$ couples the $a$ and $b$ spins at each site, but no coupling is set between neighboring $b$ spins; this picture yields a comb structure. 
Moreover, we shall see that this basis is convenient for demonstrating and characterizing the continuum limit of this fuzzy model, due to its ability to build on existing knowledge of the Heisenberg chain.

The spectrum of the spin-$1/2$ Heisenberg chain is well-known, so let us recall its relevant characteristics here for studying the low-lying spectrum of $\mathcal{H}$ as $g^2\to0$. 
In the ferromagnetic case [the ``$-$" sign in \eq{eq:fuzzysphere}], 
the ground state has spins pointing in the same direction. One-particle states (magnons) have a quadratic dispersion relation 
$E(p) \sim p^2$ ($p=0,2\pi/\mathsf{L}, \ldots$). 
In the antiferromagnetic case [the ``$+$" sign in \eq{eq:fuzzysphere}], the ground state is much more complicated; the ground state is not formed simply by alternate spins, but it can still be found by the Bethe ansatz~\cite{bethe}. 
(For a more contemporary introduction, see, e.g., Refs.~\cite{10.1063/1.4822511,10.1063/1.168740,fradkin_2013}.) 
Assuming $\mathsf{N}$ is even, the ground state then has zero total spin ($s=0$), and it has components in all states of the computational basis with vanishing total $S_z=0$. 
The excitations, in the case of half-integer spin, are gapless \cite{Haldane:1983ru} with a linear dispersion relation $E(p) \sim p$ (spinons). 
Even-$\mathsf{N}$ lattices can have only integer-spin states, and so the first excited states are spinon pairs with $s=1$.

As we will now argue, the features above strongly suggest that the model has a critical point as $g^2 \to 0$. First, since the kinetic term acts on both the left and the right of each site, it breaks the degenerate eigenspace of $\mcal V$ associated with each $\ket n$, for any $g^2 > 0$. In particular, in the ground $\mcal V$-eigenspace ($n=0$) a gap exists between a singlet ground state and a triplet of first excited states. Consequently, for any \emph{finite} lattice size, the gap must vanish as $g^2 \to 0$, since the spectrum of $\mathsf{a}\mathcal{H}$ in Eq.~\eqref{eq:fuzzysphere} reduces to that of $\mathcal{V}/g^2$ 
in that limit. 
Moreover, the degenerate spaces have dimension $\geq2^\mathsf{N}$, implying the number of states merging together as $g^2\to 0$ is exponential in the volume, as desired of any field theory. 
In strictly infinite volume, on the other hand, we cannot prove gaplessness as $g^2 \to 0$. That the infinite-volume theory indeed becomes gapless in the limit was demonstrated numerically in Ref.~\cite{Alexandru:2022son} using the density matrix renormalization group (DRMG) algorithm.\footnote{The neighbor term in the $\ell_\mrm{max}$ truncation is gapped in any finite volume and therefore does not have an exponentially degenerate ground state. Together these facts may explain the failure of that model to have a critical point~\cite{Alexandru:2022son}.}

Alone, the existence of a critical point argued above does not guarantee that one obtains the asymptotically free $\sigma$-model in the continuum limit. 
To that end, we can gain more insight by studying the small $g^2$ limit in perturbation theory.
Since the ground state of $\mathcal{V}$ is highly degenerate, we adopt the effective Hamiltonian approach to degenerate perturbation theory \cite{Sakurai:1167961}. 

\begin{figure}[tbp]
%\begin{minipage}{.45\textwidth}
    %\centering
    \includegraphics[scale=0.21]{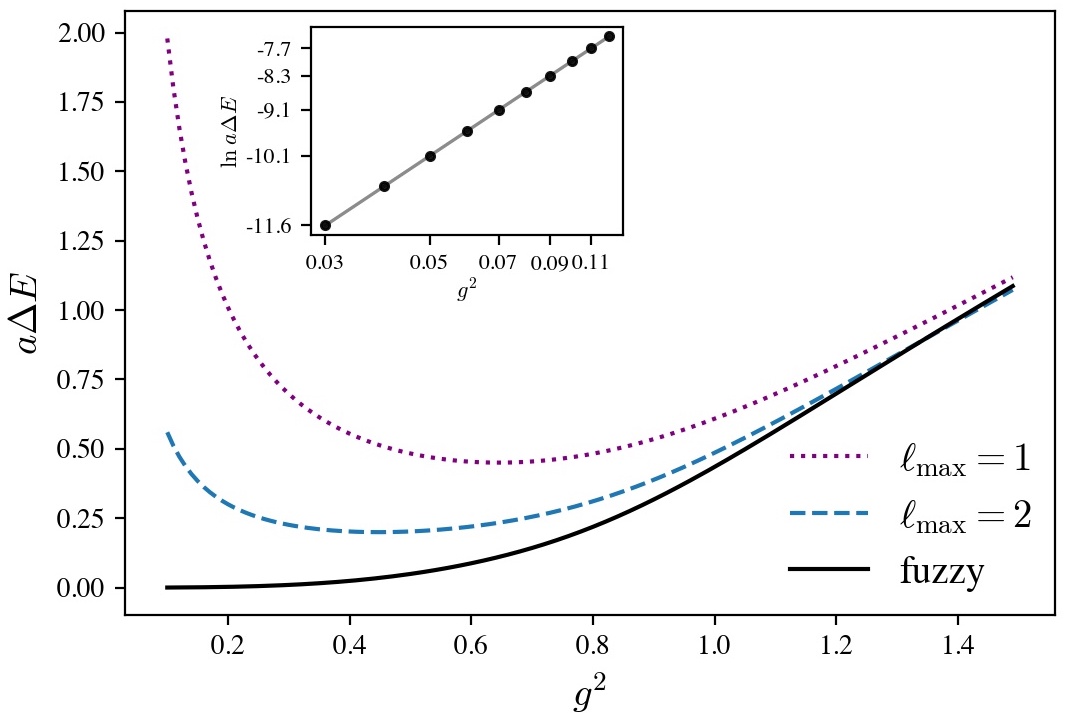}
    \caption{Comparison of the energy gaps in the fuzzy $\sigma$-model with those of the $\ell_\mrm{max}$ truncation strategy for $\ell_\mrm{max}=1,2$. The chain length is $\mathsf N = 4$ in each case. The inset displays the small-$g^2$ behavior of the fuzzy gap on a log-log scale; the slope of the line is 3.
    }
    \label{fig:sigma-gap-plot}%
%\end{minipage}%
\end{figure}

The first-order term $\mathsf{a} \mcal H^{(1)}_\mrm{eff}$ is simply the kinetic operator restricted to the ground subspace, $\{\ket{\Psi_\alpha}=\ket{0}\otimes\ket{\chi_\alpha}\}$ with $\alpha=1,\ldots,2^\mathsf{N}$,
\begin{align}
\mathcal{K}_{\alpha\beta}  & = \eta g^2 \bra{\Psi_\alpha} \sum_x \mathcal{T}(x)^2  \ket{\Psi_\beta} \\
& = \eta g^2 A_1 \delta_{\alpha \beta} - \; \eta g^2 A_2 \sum_{x,k} 
\underbrace{ \bra{0} T_k(x) \ket{0}
}_{=0} \bra{\chi_\alpha} T_k(x) \ket{\chi_\beta}, \nonumber
\end{align}
where we use the fact that $\mathcal{T}^2=[T_k, [T_k, \bullet]] = T^2 \bullet + \bullet T^2 - 2 T_k \bullet T_k$ and $T^2 \propto \bbid$, and $A_{1,2}$ are constants. 
We can see that the second term above must vanish because the Heisenberg-chain ground state is rotationally invariant. 
Since $(\mcal H_\mrm{eff}^{(1)})_{\alpha\beta} = \mcal K_{\alpha\beta} \propto \delta_{\alpha\beta}$, there is no first-order contribution to the gap $\Delta E = E_1-E_0$ in the degenerate subspace of $\ket 0$. Therefore, we expect the gap to scale as $\mathsf{a} \Delta E \sim (1/g^2) (A g^4+B g^8 + \cdots) \sim g^6$, with $A=0$.\footnote{
Since $\eta(g^2)$ converges to a positive constant as $g^2\to0$~\cite{Alexandru:2022son}, we denote this limit simply as $\eta$, with any correction resulting in a high-order function in $g^2$ that we may safely neglect, if we are only concerned with the lowest-order correction in $g^2$. 
} 
We confirm this behavior numerically by exact diagonalization of \eq{eq:fuzzysphere} in small volumes; see Fig. \ref{fig:sigma-gap-plot}. 

\begin{figure}[tbp]
    \includegraphics[scale=0.21]{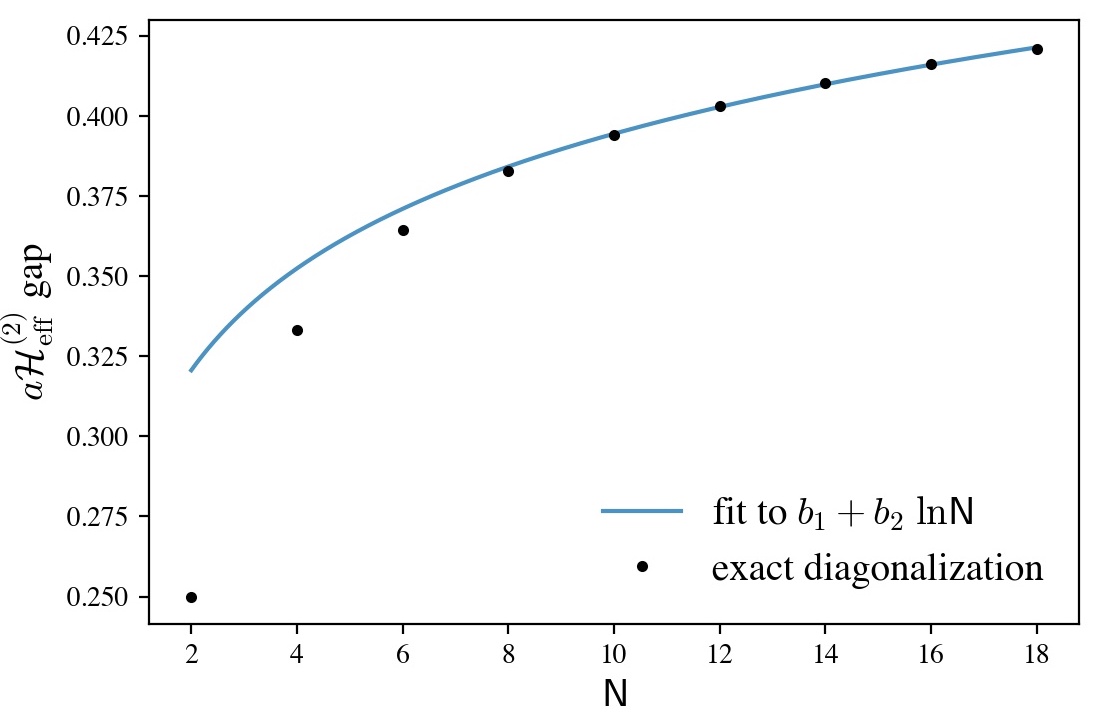}
    \caption{Behavior of the perturbative coefficient of $g^6$ in $\mathsf{a} \Delta E$ for the fuzzy $\sigma$-model, as determined by the gap in $\mcal H_\mrm{eff}^{(2)}$, \eq{Heff}, for chains up to size $\mathsf N=18$. The blue curve is a fit to $b_1+b_2\log \mathsf N$, indicating a logarithmic divergence in infinite volume.
    }
    \label{fig:log-N-plot}
\end{figure}

Although this perturbative expansion is valid in small enough boxes, $\mathsf{L}=\mathsf{a}\mathsf{N} \ll 1/M$, it breaks down in larger volumes, $\mathsf{L}=\mathsf{a}\mathsf{N} \gg 1/M$, due to an infrared divergence of the $B$ coefficient in $\mathsf{a}\Delta E$. 
We have observed this behavior numerically by directly computing $B$ via exact diagonalization up to $\mathsf{N}=18$ (see Fig.~\ref{fig:log-N-plot}), but we can also understand it analytically.
The second-order contribution to the effective Hamiltonian in the ground subspace of $\mathcal{V}$ is determined by the matrix\footnote{The perturbed \textit{eigenstates} are found by choosing the unperturbed kets to be eigenstates of $\mcal H_\mrm{eff}^{(2)}$, instead of the leading matrix $\mcal K$, in contrast to standard degenerate perturbation theory.
}
\
\begin{align} \label{Heff}
\mathsf{a}(\mathcal{H}&^{(2)}_\mathrm{eff})_{\alpha\beta} = \bra{\Psi_\alpha} \mcal K P_1 \big[ v_0 P_1 - P_1\mcal VP_1]^{-1} P_1 \mcal K \ket{\Psi_\beta} \nn\\
& =  2\sum_{m\geq 1} \sum_{x,y} 
\frac{\eta g^6}{v_m-v_0}
\bra{0} T_i(x)  \ket{m}  \bra{m}  T_j(y)  \ket{0}   \nonumber\\
& \qquad\qquad
\times
\bra{\chi_\alpha}  T_i(x)  T_j(y)  \ket{\chi_\beta}, 
\end{align}
where 
$P_1$ is a projector to the excited subspaces $\{ \ket m \otimes \ket \chi : m \geq 1 \}$  of $\mathcal{V}$, and $v_0, v_m$ are the respective eigenvalues of $\ket 0, \ket m$. The $m$-sum includes the low-lying two-spinon states with $v_m-v_0\sim p$ ($p\approx 2\pi/\mathsf{L}, \ldots$, at large $\mathsf{L}$). Note that the matrix elements 
 $\bra{0}T_i(x)\ket{m}$ do not vanish for states $\ket m$ with total spin 1. 
Therefore, for large $\mathsf{L}$, unless those matrix elements vanish at small $p$, the sum over $m$ will diverge logarithmically:
\beq
\mathsf{a}\Delta E \sim \int_{2\pi/\mathsf{L}}^{\pi/\mathsf{a}} \frac{\mathrm{d}p}{p} \sim \ln (\mathsf{L}/\mathsf{a}),
\eeq
indicating the breakdown of perturbation theory in large volumes.

The features described above, namely, the existence of a critical point with an exponentially degenerate Hilbert space, and the breakdown of perturbation theory at large distances, are hallmarks of asymptotically free continuum field theories.

\subsection{Gauge theory}
\label{sec:continuum-lgt}

Much of the reasoning above carries over to the fuzzy gauge theory, with a Hamiltonian given in \eq{eq:ham-generic}.
Again, the potential (plaquette) operator acts only from the left on fuzzy states, and the eigenstates factorize as before: $\ket{\Psi_n}=\ket{n} \otimes\ket{\chi}$ with potential energy $v_n$. 
Here the $\ket n$ are eigenstates of a plaquette operator acting on a four-dimensional one-link Hilbert space (rather than $4\times 4$ matrix states). 
In fact, this ``reduced'' plaquette operator is just the simplest example of a ``gauge magnet'' as proposed by Orland and Rohrlich in Ref.~\cite{Orland:1989st}.
In a spin-wave analysis, the pure-plaquette case of the gauge magnet was argued to be gapless with a nonrelativistic dispersion relation. 
However, since the eigenstates are independent of the overall coupling of the plaquette term, so is the correlation length (in lattice units). 
Therefore, the coupling cannot be tuned such that $\xi/\mathsf{a}\rightarrow \infty$ (continuum limit) while $ \xi \rightarrow$ constant, as is necessary to reproduce the physics of an asymptotically free gauge theory. 
An equivalent argument is to point out that the usual mechanism of ``dimensional transmutation" can occur only if there is a coupling constant to be tuned to alter the eigenstates.\footnote{
The gauge magnet of Ref.~\cite{Orland:1989st} was proposed as a candidate for a finite-dimensional model in the same universality class as Yang-Mills gauge theory. 
This model allows for the inclusion of a kinetic term which acts on each four-dimensional one-link Hilbert space via multiplication with $\gamma_5$. 
We have exactly diagonalized this model in the one-plaquette (with open boundaries) and four-plaquette (with periodic boundaries) systems. 
The behavior for small $g^2$ is qualitatively similar to the $j_\mrm{max}$ qubitizations. 
Still, a gapless point at a nonzero $g^2$ exists in the four-plaquette system (though not in the one-plaquette system). 
It is not clear if the number of degenerate states is exponential in the volume, unlike the fuzzy model. 
In any case, this model warrants further study by nonperturbative means. See also Ref.~\cite{Orland:2013}. 
}

\begin{figure}[tbp]
    %\centering
    \includegraphics[scale=0.55]{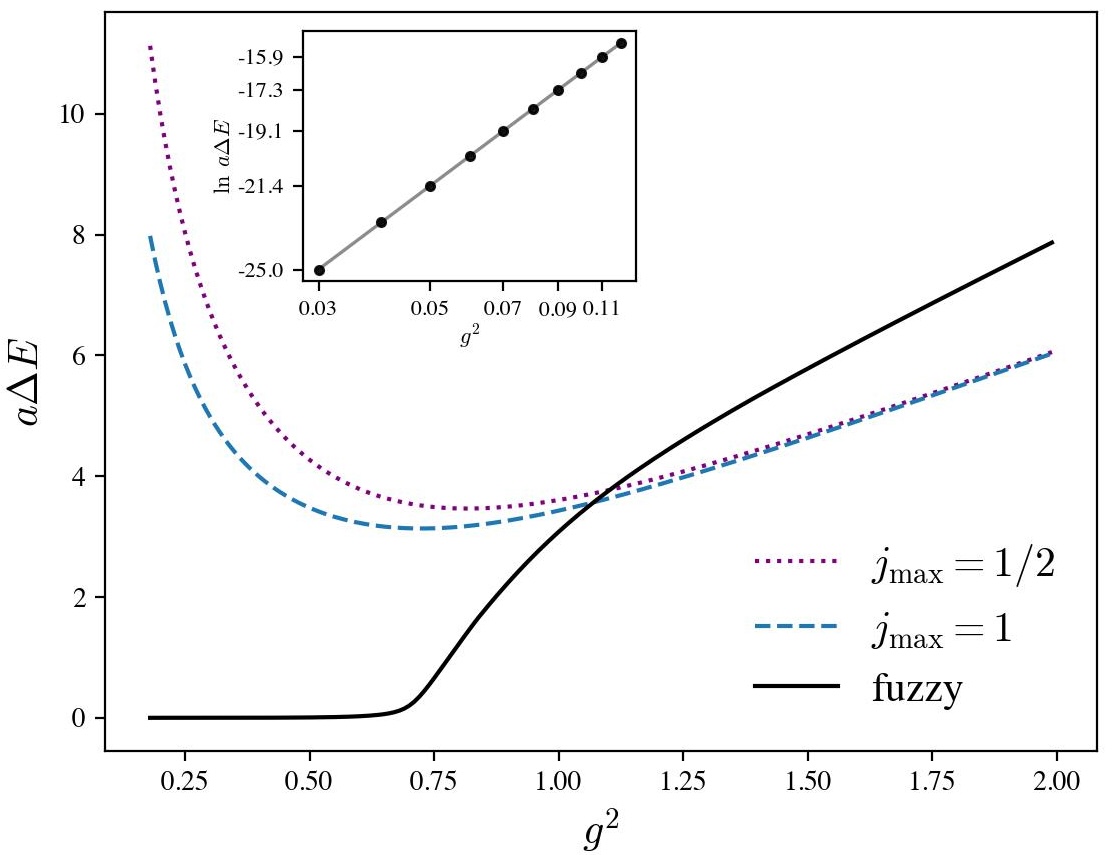}
    \caption{Energy gaps in the \textit{physical} subspace of qubitized $SU(2)$ gauge theories for a single-plaquette system. 
    We compare the fuzzy model (with kinetic term $\mcal K_2$) to the angular momentum $j_\mathrm{max}$ truncation models discussed in the Introduction. 
    In each case, as $g^2\to 0$, the behavior is dominated by the plaquette operator $\square$. 
    For the $j_\mrm{max}$ models the finite gap of $\square$ yields a diverging energy gap, while for the fuzzy model the degeneracy of $\square$ ensures gaplessness. 
    The inset plots the fuzzy gap versus $g^2$ on a log-log scale for small $g^2$; a fit to a straight line is in gray, with a slope of $7$.
    }
    \label{fig:gap-results}
\end{figure}

Due to the factorization property outlined above, each eigenvalue $v_n$ of the \emph{fuzzy} plaquette operator is at least $4^\mathsf{N}$-fold degenerate (for now, including unphysical states), where $\mathsf{N}$ is the number of links. 
{As we detail below, the degeneracy of the ground state is lifted by a suitable choice of kinetic term.
In \fig{fig:gap-results} we confirm this behavior by exact diagonalization in a small volume, where also the gaps are compared with the two simplest angular momentum $j_\mathrm{max}$ truncations (described in Sec.~\ref{sec:intro}).}
As $g^2 \to 0$, the low-lying states merge into the degenerate ground subspace of the fuzzy plaquette operator, and the number of such states grows exponentially with the volume, which {strongly suggests} the existence of a critical point suitable for continuum physics. 

The choice of kinetic terms $\mathcal{K}_1$, $\mathcal{K}_2$, or some combination thereof, is unclear. 
Both operators are positive-semidefinite, as in the standard Kogut-Susskind Hamiltonian. 
While $\mathcal{K}_1$ resembles the kinetic term of the Kogut-Susskind Hamiltonian, since it is the quadratic Casimir on the fuzzy Hilbert space, it does not completely lift the degeneracy of the ground state in the one-plaquette system; a twofold degeneracy in the ground state subspace, not present in continuum gauge theories, remains. 
Meanwhile, $\mathcal{K}_2$ \textit{does} completely lift this degeneracy, but it is not as directly associated with the Casimir operator of the original gauge theory. 
Only a deeper insight into the dynamics of fuzzy gauge theories can settle which\textemdash if any\textemdash kinetic terms lead to the proper continuum limit. 

Furthermore, without a complete understanding of the dynamics of the Orland-Rohrlich gauge magnet, we cannot provide an analytic argument for the breakdown of perturbation theory with increasing volume, nor can the fuzzy gauge theory be exactly diagonalized beyond the one-plaquette system on current hardware. 
Still, we have observed in the one-plaquette system that the leading order term in the effective Hamiltonian of the degenerate ground state subspace must scale as $\mathsf{a}\Delta E \sim g^{14}$, as depicted in \fig{fig:gap-results}, which corresponds to fourth-order perturbation theory in $g^4$. 
This high-order first nonvanishing correction  suggests impracticality and perhaps even lack of efficacy of perturbation theory in studying this theory (i.e., if nonperturbative in nature).
Likewise, if the spin-wave analysis of the Orland-Rohrlich model holds, where the gauge magnet is predicted to be gapless, then it is possible for the perturbative coefficients of the fuzzy gauge theory to have infrared divergences in a manner similar to the $\sigma$-model analysis of Sec.~\ref{sec:continuum-sigma}.

Before we conclude our discussion, let us justify a subtle problem of local gauge theories not present in the theories with only global symmetries: ensuring the exponential growth of the {\it physical} Hilbert subspace with volume. 
We implicitly claimed earlier that the growth in the ground state degeneracy survives the restriction to the physical subspace; we argue for this point as follows. 
First note that in the product basis, the all-left-multiplying plaquette operator of our fuzzy $SU(2)$ model is simply a tensor product of the gauge magnet plaquette operator and an identity operator of the same dimension, as remarked earlier. 
Let $\ket n$ be a gauge-invariant eigenstate of the Orland-Rohrlich total plaquette operator with energy $v_n$. 
Then, the degenerate $v_n$-eigenspace of the \textit{fuzzy} plaquette operator includes 
states that, in the $\mrm{E}_{ab}$ product basis, take the form
\BE
\ket {\Psi_n} = \ket n \otimes \ket{\bo \chi},
\EE
where $\ket{\bo \chi}$ is a $4^\mathsf{N}$-dimensional state.
The action of the gauge generator $\mathcal{G}_a(x)$ 
in the fuzzy theory acts on this basis as
\BE
\mathcal{G}_a(x) \ket{\Psi_n} = - \ket n \otimes G_a(x) \ket{\bo \chi},
\EE
since $\ket n$ is gauge-invariant, $G_a(x)\ket n = 0$; cf.~Eq.~\eqref{eq:fuzzy-phys}. 
Therefore, the \textit{physical} fuzzy states with plaquette energy $v_n$ include all states satisfying $G_a(x) \ket{\bo \chi} = 0$, and the set of such states is the entire physical subspace of the Orland-Rohrlich gauge magnet, with dimension $D_\mrm{OR}$. 
We note that there also exist physical states in the fuzzy model that cannot be written as the tensor product of two invariant gauge-magnet states. 
Thus the dimension $D_f$ of the physical fuzzy Hilbert space satisfies
$D_{f} \geq (D_\mrm{OR})^2$, 
and the dimension $D_{f0}$ of the physical fuzzy degenerate ground subspace satisfies $D_{f0} \geq D_\mrm{OR}$. 

It then follows that if the Orland-Rohrlich model has a physical Hilbert space that is exponential in the volume, 
then the (physical) degenerate ground state subspace of the fuzzy plaquette operator will be exponentially large as well, ensuring that the number of physical states which merge in the $g^2 \to 0$ limit is exponential. 
We have verified numerically, by computing the nullity of $\sum_{x,a}G_a(x)^2$, that the Orland-Rohrlich model has two physical states in the one-plaquette system and 50 physical states in the four-plaquette system. 
More generally, based upon the diagrammatic arguments of Ref.~\cite{Orland:2013} (i.e., observing that there is at least one extra degree of freedom per site in a diagram representing a global physical state on the lattice), we expect the physical Hilbert space in the fuzzy $SU(2)$ gauge theory to grow at least exponentially in the volume, for all spatial dimensions $d$.

\section{Quantum Circuitry}
\label{sec:quantum-sim}

In this section, we explain how we construct a quantum circuit performing the Hamiltonian time evolution, working explicitly with the case of $SU(2)$. 
To this end, we compute the Hamiltonian in the computational basis given by single-entry matrices, $\mrm{E}_{\alpha\beta}$ ($\alpha,\beta=0,\ldots,3$). 
If we flatten the indices as $\alpha\beta\mapsto A$ with $A=1,\ldots,16$, then the operator $\mathcal O$ represented in this computational basis has elements 
\BE
[\mathcal O]_{A,B} := \mrm{tr} \big[ \mrm{E}_A^\dag \, \mathcal O \, \mrm{E}_B \big].
\EE 
With a local Hilbert space dimension of $16=2^4$, we can neatly encode these operators by their action on four qubits $\ket{q_1,q_2,q_3,q_4}$, where $q_i=0,1$ encode $A-1$ in a binary representation. 
We build the link operators out of gamma matrices, which are given by
\begin{align}
    \Gamma_\alpha \doteq \frac{1}{2} \gamma_\alpha \otimes \bbid_4 = \frac{1}{2} \begin{pmatrix}+XI\\-YX\\-YY\\-YZ\end{pmatrix} \otimes II
    \label{eq:fuzzy-gamma_qubits}
\end{align}
[cf.~Eq.~\eqref{eq:d4}] to compose the $\f U$ as in Eq.~\eqref{eq:fU-gamma}, as well as 
\begin{align}
\Gamma_5\doteq \frac{1}{2}\gamma_5\otimes\bbid_4 = -\frac{1}{2} ZIII.
\end{align}
The local operators relevant for the kinetic terms, Eqs.~\eqref{kinetic-1} and \eqref{eq:kin-u}, are given by
\begin{align}
    (\mathcal{T}^{L})^2 + (\mathcal{T}^{R})^2 
    \doteq &-I\,XI\,X+I\,YI\,Y-I\,Z\,I\,Z
    \nonumber \\
    &-ZXZX+ZYZY-ZZZZ
    \label{eq:fuzzy-kinetic_qubits}
\end{align}
and
\begin{align}
    - \fU_{ij} & \bullet \fU_{ij}^\dag - \fU_{ij}^\dag \bullet \fU_{ij} \nonumber\\
    \doteq& -XIXI + YXYX- YYYY+ YZYZ.
\end{align}
%(up to a shift proportional to identity $\bbid_{16}\doteq IIII$) 
These are relatively easy to include in quantum simulations.

\begin{figure}[tbp]
    \centering
    \includegraphics[width=0.475\textwidth]{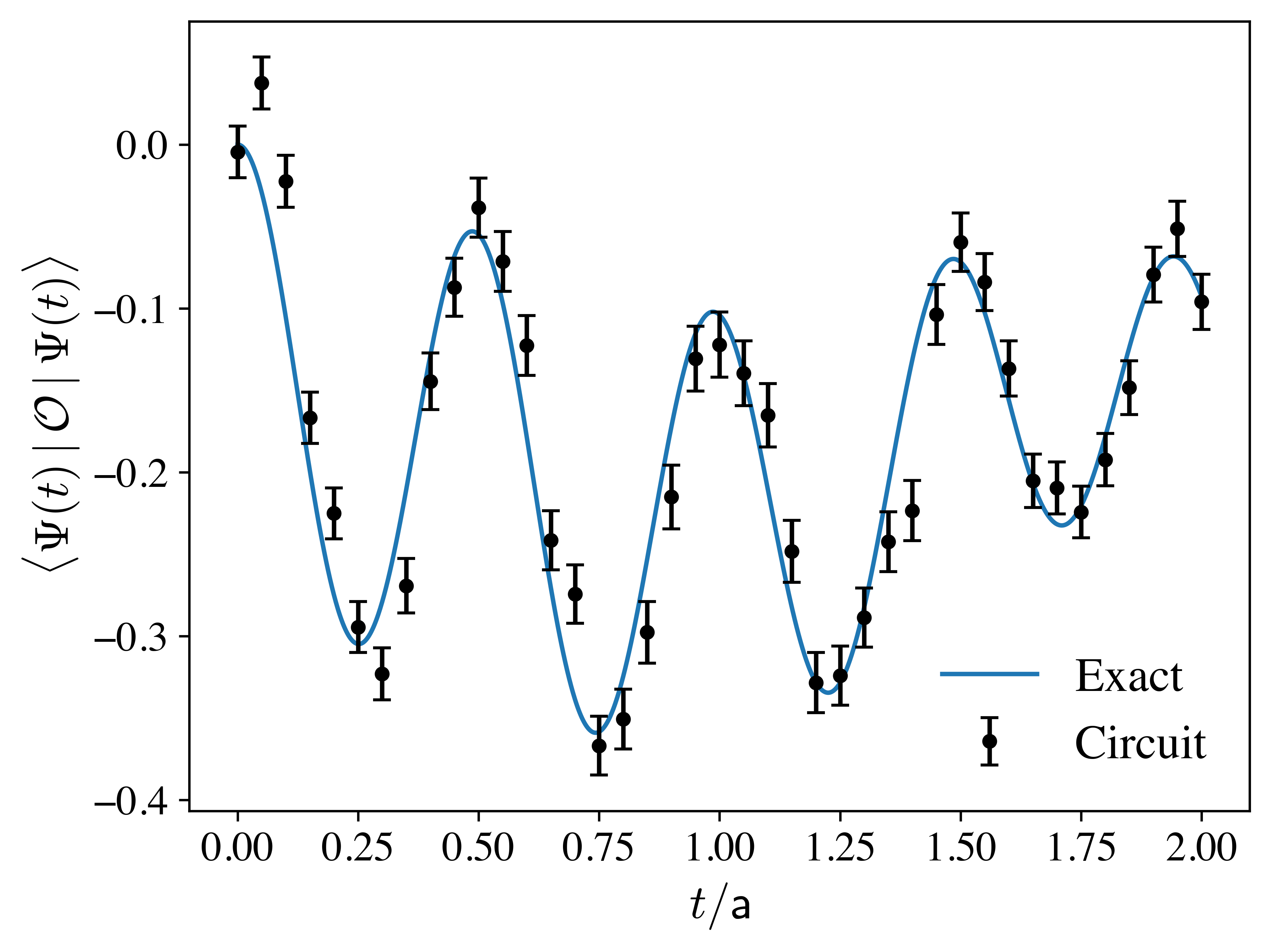}
    \caption{
    Expectation value of the plaquette operator $\mathcal{O}=\square$ with respect to the evolved state $\ket{\Psi(t)} = \exp(-\mathrm{i}t\mathcal{H}) \ket{\Psi(0)}$, choosing a gauge-invariant initial state $\Psi(0)\doteq\bbid_4\,^{\otimes4}{/16}$ and kinetic term $\mathcal{K}_2$ {and $g^2=1$} for example. 
    The solid line is obtained by 
    numerically calculating $\Psi(t)$. 
    The data points shown are obtained by repeatedly applying the quantum circuit of $Q\approx\exp(-\mathrm{i}\delta \mathcal{H})$, constructed as described in Appendix~\ref{sec:compile} and using the prescription in Ref.~\cite{Murairi:2022zdg}. 
    Uncertainty bars on each data point show the 
    shot noise of 4000 measurements at each time value. 
    Time steps of size {$\delta=0.04\,\mathsf{a}$} %$\delta=0.0025 \,\mathsf{a}$ 
    were taken. %, while the results of only every other time step are shown for clarity in the plot. 
    Additionally, the state at each time $\Psi(t)$ was prepared by applying $t/\delta$ iterations of $Q$ to $\Psi(0)$, independently of the results of $\Psi(t-\delta)$, leaving measurements at different times uncorrelated. 
    }
    \label{fig:qsim-results}
\end{figure}

With these local operators encoded on qubits in hand, we may straightforwardly construct quantum circuits to simulate the real-time evolution of the Hamiltonian of Eq.~\eqref{eq:ham-generic} with $\mathcal{K}=\mathcal{K}_2$ of Eq.~\eqref{eq:kin-u}. 
For details elaborating how we compile such circuits, please refer to Appendix~\ref{sec:compile}. 
In summary, we find that 236 {\small CNOT} gates {and 88 rotation gates} are involved in simulating both the kinetic and potential terms of a Hamiltonian for an individual plaquette. 
In addition, we depict a classical simulation of the quantum circuit for a one-plaquette system 
in Fig.~\ref{fig:qsim-results}. 
{[There, the initial state of $\Psi(0)=\bbid_4\,^{\otimes4}/16$ was prepared with two {\small CNOT} gates and two Hadamard gates per link.]} 

The relatively low gate count in a circuit simulating one plaquette may be surprising when compared with the counts found in Ref.~\cite{Murairi:2022zdg}: 180 and $\sim$16000 {\small CNOT} gates demanded by the models of Orland and Rohrlich~\cite{Orland:1989st} and Horn~\cite{HORN1981149}, respectively. 
In those cases, the operations on a given link were four- and five-dimensional, respectively, and gate costs were dominated by the number of Pauli operators involved in the plaquette term of each model. 
We also find that the large resource costs seen for the Horn model are similar for the angular momentum $j_\mathrm{max}$ truncation strategy compared earlier. 
Moreover, one could see for Hamiltonians approximating $\sigma$-models that gradually lifting a truncation in angular momentum cutoffs would necessitate more qubits to encode the models and consequently involve much greater gate costs to simulate the many new Pauli operators involved~\cite{Alexandru:2022son}, so we expect this trend to apply for the $j_\mrm{max}$ truncation as well. 

As a result, one might then be apprehensive that a model encoding an individual link's degrees of freedom on four qubits instead of two or three would entail much greater complexity to simulate. 
However, note that the fuzzy link operators as represented in Eq.~\eqref{eq:fuzzy-gamma_qubits} are proportional to those of the Orland-Rohrlich model. 
Therefore, the cost in quantum gates demanded to simulate our fuzzy plaquette of Eq.~\eqref{eq:fuzzy-plaquettes} is exactly the same as for the plaquette acting on only two-qubits per link. 
The only additional costs from our proposed fuzzy gauge model come from the kinetic term in Eq.~\eqref{eq:fuzzy-kinetic_qubits}, which is now somewhat more involved. 
While Ref.~\cite{Orland:1989st} proposed a kinetic term proportional to $\gamma_5$ on each link, which can be simulated with a single rotation gate, 
our one-link fuzzy kinetic terms involve a handful of {\small CNOT} gates.

\section{Conclusions}
\label{sec:conc}

In the above we have proposed a novel procedure to render a bosonic gauge theory finite-dimensional. 
We hope that this scheme is a helpful step in the ongoing effort to develop scalable quantum computation methods that can be applied to nuclear physics problems. In fact, it is likely that such a scheme will be \textit{necessary}, given that the removal of a truncation, such as in the $j_\mrm{max}$ truncation strategy, is prohibitively expensive in a quantum circuit.
As noted before, the proposed Hamiltonian is relatively straightforward to simulate, allowing one to work in only the number of spacetime dimensions required by the theory and sidestepping the need for a truncation or other regularization to be removed. 

A central motivation for our work was the search for a theory with a finite-dimensional Hilbert space  (per volume) and in the same universality class as non-Abelian gauge theories. 
We proposed a theory that exhibits several features (listed below) that are plausibly necessary toward that end. Unfortunately, a nonperturbative numerical calculation providing clear evidence for the correct continuum limit has not yet been possible. 
Preliminary attempts to simulate the Hamiltonian of the fuzzy gauge theory via Monte Carlo suggest, as in many truncated theories, the presence of a severe sign problem.
In the simpler case of the $(1+1)$-dimensional qubitized $\sigma$-model, tensor network methods \cite{Alexandru:2021xkf,Alexandru:2022son} and Monte Carlo methods specific to $1+1$ dimensions 
\cite{Singh:2019uwdsuu,Bhattacharya:2020gpm} were used to provide such evidence. 
Neither method is available for $(3+1)$-dimensional gauge theories, though tensor network methods may be feasible in $2+1$ dimensions. Alternatively, if a renormalization group argument valid at weak coupling could be devised, it would shed light on this question; we are actively pursuing such a strategy.

{
An appealing aspect of the fuzzy strategy\textemdash for both $\sigma$-models and gauge theories\textemdash is the formal similarity with 
the corresponding original lattice theories; in particular, states of the fuzzy models can be thought of as ``functions'' of the noncommutative $\f U_{ij}$, the symmetry generators satisfy the Leibniz rule on any state, and the $\f U_{ij}$ satisfy an analog of the classical special-unitarity constraints. 
Moreover, we have identified throughout the paper various concrete properties that we {expect} any qubitization of non-Abelian gauge theory %must 
{should} satisfy. 
Namely, the qubitization should possess

\begin{enumerate}[label=(\Alph*)]
\item the same gauge symmetry as the original theory [i.e., $SU(N)$], 

\item the same global symmetries as the original theory (i.e., center $\mathbb{Z}_N$ symmetry), 

\item a critical point at fixed, finite local Hilbert space dimension, 

\item the desired spacetime symmetry (i.e., Lorentz covariance) in the critical limit,

\item a {gauge-invariant} Hilbert {sub}space that grows exponentially with the volume,

\item a nonperturbative mass gap consistent with asymptotic freedom, and

\item a small local Hilbert space, %convenient for encoding on quantum hardware. 
for easy qubit-embedding. 
\end{enumerate}
% ABCDEFG -> ABDF CEG
The conventional wisdom is that merely possessing the right symmetries, (A) and (B), would suffice to place a model in the correct universality class. 
However, a lesson of searches for qubitizations of the $\sigma$-model is that more properties are required~\cite{Alexandru:2022son}, as emphasized by the list above. {We emphasize that the list above is not necessarily a minimally redundant list; for example, it is possible that conditions (A)-(D) provide the only necessary criteria. However, in the absence of complete knowledge of the dynamics of our theory, we include possibly redundant criteria to serve as indicators that the model is on track.}

Let us now collect what evidence we have produced for these conditions in fuzzy gauge theory. 
By its definition in Sec.~\ref{sec:fuzzy-algebra}, the fuzzy gauge theory satisfies condition (A), and we demonstrated that it 
also satisfies (B). 
In Sec.~\ref{sec:continuum}, we first reviewed how the fuzzy $\sigma$-model satisfies conditions analogous to (A), (B), and (E) before providing new evidence in favor of conditions (C) and (F); while these properties were not completely proven here, 
they were previously demonstrated nonperturbatively using DMRG~\cite{Alexandru:2021xkf,Alexandru:2022son}. 
We then showed evidence that conditions (C), (E)-(G) are again satisfied by the fuzzy gauge theory, but we 
cannot produce as much 
evidence for condition (F) in fuzzy gauge theory as in the fuzzy $\sigma$-model. 
We cannot yet assert condition (D) with any confidence. 
Condition (G), though not strictly necessary, is highly desirable for efficient simulation, and we have demonstrated the reasonable circuit depth of the fuzzy qubitization in Sec.~\ref{sec:quantum-sim}. 
}

Over the decades much effort has been put into understanding gauge dynamics from the Hamiltonian point of view, starting with Ref.~\cite{Kogut-Susskind} in the 1970s.
It is possible that a fuzzy gauge theory, by virtue of its simplicity, while still conjectured to be equivalent to continuum gauge theory, might be used to advance this program. 
Historically, many-body physics had several breakthroughs based on wave function ans\"atze, but the same has not yet occurred in field theory (see, for instance,  Refs.~\cite{Greensite:1979,Greensite:2007,Karabali:2009,Smith:1992,Leigh:2006vg,Raychowdhury:2018} for work in this direction). 
Perhaps {the fuzzy approach that we have explored here} will help in this effort.

\begin{acknowledgments}
We thank Thomas Cohen, Shailesh Chandrasekharan, Aleksey Cherman, and Maneesha Pradeep for helpful conversations. 
This work was supported in part by the U.S. Department of Energy, Office of Nuclear Physics under Award No.~DE-SC0021143, No.~DE‐FG02‐93ER40762, and No.~DE-FG02-95ER40907. 
The numerical results were produced in part with resources provided by the High Performance Computing Cluster at The George Washington University, Research Technology Services. 
This research uses resources of National Energy Research Scientific Computing Center (NERSC), a U.S. Department of Energy Office of Science User Facility located at the Lawrence Berkeley National Laboratory.
\end{acknowledgments}

\appendix

\section{A generalization to \texorpdfstring{$U(N)$}{Lg} gauge groups}
\label{sec:fun-in-the-un}

As mentioned in Sec.~\ref{sec:fuzzy-algebra}, one can come up with explicit matrices to satisfy fuzzy unitarity constraints. 
Here, we provide an example of these matrices here for a generalization to $U(N)$ 
and discuss obstacles to efficiently reduce the theory to avoid a potentially undesired extra $U(1)$ gauge symmetry in this example.

For the case of fuzzy unitary matrices for large gauge groups, 
we propose a subtly different parametrization of matrices in $U(N)\ni U=x_0\openone_N+x_a\lambda_a$, where $a=1,\ldots,N^2-1$ and $\lambda_a$ are generalized Gell-Mann matrices. 
In this case, we allow $(x_0,x_a)\in\mathbb{C}^{N^2}$, as the complex span of $\openone_N$ and $\lambda_a$ includes all unitary matrices. 
Correspondingly, the operators to which we will promote these parameters are not necessarily Hermitian. 

Based upon a complex parametrization of the $U(N)$ manifold, 
we could propose new operator-valued parameters form a link operator 
\begin{align}
    \f{U}_{ij} = \delta_{ij} X_0 + (\lambda_a)_{ij} X_a.
    \label{eq:fuzzy-link-N} 
\end{align}
Given this parametrization of $\f{U}$, 
one can show that a slightly different operator constraint for fuzzy unitarity,\footnote{Notably, this constraint is {\it not} gauge-invariant for $N\geq2$. That is, non-Abelian gauge transformations of matrices satisfying these anticommutation relations do not result in more fuzzy unitary matrices. Nevertheless, gauge-invariant states and Hamiltonian operators can be constructed from these matrices.}
\BE
\frac{1}{2} \{ \fU, \fU^\dag \}_{ij} =\delta_{ij} \bbid,
\label{eq:fuzzy-unitarity-II}
\EE
is equivalent to $N^2$ constraints
\begin{equation}
    N\{X_0,X_0^\dagger\} + 2\{X_a,X_a^\dagger\} = 2N\openone, \label{eq:fuzz-ellipsoid} 
\end{equation}
\begin{equation}
    \{X_a,X_0^\dagger\}+\{X_0,X_a^\dagger\} + h_{abc}(X_bX_c^\dagger + X_b^\dagger X_c)=\mathbbm{0}, \label{eq:fuzz-vector}
\end{equation}
where $h_{abc}=d_{abc}+\mathrm{i}f_{abc}$ with totally symmetric and antisymmetric structure constants $d$ and $f$, respectively. 
One finds that the above equations are satisfied by 
\begin{align}
    X_0 = \frac{1}{r_0} &\sigma_+ \otimes \openone_N \\
    X_a = \frac{1}{r_a} &\sigma_+ \otimes \lambda_a,
\end{align}
where $\sigma_+=(\sigma_x+\mathrm{i}\sigma_y)/2$, $r_0^2=N^2/2$, and $r_a=(2/N)r_0$. 
Fascinatingly, one can then see for the matrix elements of $\f{U}$ that
\begin{align}
    \f{U}_{ij} &= \sqrt{2}\,\sigma_+\otimes \mathrm{E}_{ji},
\end{align}
using the Fierz identity of $\mathrm{SU}(N)$ for $\lambda_a\otimes \lambda_a$. 
Here, we have defined the single entry matrix $\mathrm{E}_{ij}=\me_i\me_j^{\top}$ (with unit vectors $\me_i$). 
In other words, these $\f{U}$ matrices are actually similar to those proposed in Ref.~\cite{Brower:1997ha}, albeit with different normalization and intended here for action on a matrix-valued Hilbert space.\footnote{We note that our use of the quantum link matrices does not involve an appeal to dimensional reduction, and as such will be formulated directly in the space dimension of interest.} 

Moreover, we demand that these new link operators $\f{U}$ behave as expected under gauge group transformations. 
Technically speaking, 
we find that the Kogut-Susskind relations of $\f{U}_{ij}$ of Eqs.~\eqref{eq:lU-comm} and \eqref{eq:rU-comm} are equivalently
\begin{align}
    [T^{R/L}_A,X_0] &= \pm \frac{1}{N} X_a, \\
    [T^{R}_a,X_b] &= +\frac{1}{2} \delta_{ab} X_0 + \frac{1}{2} h_{abc} X_c, \\
    [T^{L}_a,X_b] &= -\frac{1}{2} \delta_{ab} X_0 - \frac{1}{2} h_{abc}^* X_c.
\end{align}
With the natural choice of $T^{R/L}_a = (1\pm\sigma_z) \otimes \lambda_a /4$ (just as we did in Sec.~\ref{sec:fuzzy-algebra}),
we find that all of the above commutation relations are indeed satisfied.

Finally, let us point out that these $X$ operators are nilpotent with degree 2, meaning $X X' = \mathbbm{0}$ and so $\f{U}$ inherits this nilpotence. 
Nonetheless, importantly, $X^\dagger X'$ and $XX^{\prime\dagger}$ are nontrivial, indicating $\f{U}_{ij}\,(\f{U}_{kl})^\dagger$ and $(\f{U}_{ij})^\dagger\,\f{U}_{kl}$ $\neq\mathbbm{0}$, and the link operators' products with generators $T^{R/L}_a$ are also nontrivial. 
Still, this property does mean that most generalizations of the determinant for matrices over noncommutative rings still result in $\det\f{U}=\mathbbm{0}$. 
While one might desire an operator $\det\f{U}=\openone$ for a ``fuzzy'' $SU(N)$ theory, we note that this difference does not necessarily preclude a prediction of an $SU(N)$ theory by this model if one can still decouple a possible extra $U(1)$ symmetry lingering in the formulation, as we shall describe below. 
Additionally, there may be benefits to this nilpotence; certain extra gauge-invariant operators that may complicate the Hamiltonian of this model are trivialized in this way. 
For example, consider for $N=2$ the operator on two adjacent links 
\begin{align}
    \mcal O = \varepsilon_{il} \varepsilon_{kn} \,\f{U}_{ij}(1) \,\f{U}_{jk}(2) \,\f{U}_{lm}(1) \,\f{U}_{mn}(2)
    \label{eq:2-link-op}
\end{align}
with $\varepsilon=\mathrm{i}\sigma_y$, 
which is gauge-invariant as long as the Kogut-Susskind relations are satisfied (otherwise regardless of one's choice of promoting $U\mapsto\f{U}$), and yet $\f{U}_{ij}(\ell) \, \f{U}_{lm}(\ell)=\mathbbm{0}$ for either $\ell=1,2$.  
As such, this nilpotence property may help keep the form of our Hamiltonian particularly simple, involving our usual plaquette term and few other options for kinetic terms. 

In this distinct proposal for matrices $\fU$, there do exist generators of $U(1)$ gauge transformations: $T^{R/L}_0 = (1\pm\sigma_3) \otimes \bbid_N$. 
In this case, there are multiple strategies for dealing with the additional symmetry. 
The first is to explicitly break the $U(1)$ symmetry by adding to the Hamiltonian a noninvariant term. 
The simplest option would be a generalized determinant $\varepsilon_{ik} \fU_{ij} \fU_{kl} \varepsilon_{jl}$; however, this operator vanishes in the fundamental rep of $SU(2N)$. 
Reference~\cite{Brower:1997ha} suggests working with larger reps of $SU(2N)$, in which the determinant is nontrivial, but doing so here increases the fuzzy Hilbert space dimension; for $N=2$, it would be 36-dimensional. %} 
Alternatively, one could implement a $U(1)$ projector on each one-link Hilbert space.

\section{Compiling quantum circuits for a fuzzy gauge theory}
\label{sec:compile}

In this section, we detail the methods used to compile the quantum circuits for which we present simulation results in Sec.~\ref{sec:quantum-sim}. 
As a broad outline: 
We first trotterize our time evolution operator into a sequence of operators, each of which we may diagonalize relatively efficiently. 
To complete the compilation procedure, we use methods prescribed in an earlier article~\cite{Murairi:2022zdg} to determine circuits that perform the appropriate multiqubit rotations to simulate the diagonalized unitary operators produced by the earlier step. 

As is common in procedures for quantum compilation of unitary circuits $\exp(-\mrm{i}Ht)$ 
and as per the approach of Ref.~\cite{Murairi:2022zdg}, 
we choose to 
expand the Hamiltonian $H$ of Eq.~\eqref{eq:ham-generic} in the Pauli operator basis, 
\BE
H = \sum_{j} c_j P_j, 
\EE
where $P_j \in \{I, X, Y, Z\}^{\otimes n}$ (a tensor product of $n$ Pauli matrices), 
$n$ is the number of qubits chosen to represent the system on a quantum computer, 
and $c_j = \mrm{tr}[P_j H]/2^n$. 
Moreover, as in Ref.~\cite{Murairi:2022zdg}, we perform a trotterization by collecting commuting Pauli operators. 
That is, we write
\BE
\label{eq:clustering}
H = \sum_{j = 1}^{k} h_j
\EE
where Pauli operators comprising a given $h_j$ commute with one another. 
As in Refs.~\cite{Murairi:2022zdg, vandenBerg2020circuitoptimization,Tomesh:2021pns}, collecting these commuting Pauli operators can be reduced to the well-known graph coloring problem. 

In the case of a single plaquette system, we use $n = 16$, 
and decomposing its Hamiltonian in this fashion yields $88$ Pauli operators with nonzero coefficients, excluding the identity. 
Moreover, we use the DSATUR algorithm of Ref.~\cite{brelaz-paper} to 
obtain $k = 5$ commuting clusters of Pauli operators. 
Now, having reduced our problem to the simulation of commuting clusters $\{h_j\}$, 
we turn our attention to how to compile a circuit for any one such Hamiltonian $h_j$. 

Since Pauli operators comprising a cluster $h_j$ commute with each other, they can be exactly diagonalized simultaneously. 
Moreover, such a diagonalization can be performed with $\mathcal{O}(n^2)$ {\small CNOT} gates~\cite{vandenBerg2020circuitoptimization,Crawford2021efficientquantum}. 
In particular, let $V_j$ be an unitary transformation diagonalizing $h_j$. We obtain $V_j$ via the method developed in Ref.~\cite{PhysRevA.108.062414}, which guarantees shallow circuits with depths $\mathcal{O}\left(n \log_2 r \right)$ where $r\leq n$ is the number of independent generators of the commuting cluster.
As an upshot, we can then reduce $D_j = V_j h_j V^\dagger_j$, and so overall 
\begin{align}
    e^{-\mathrm{i} H \delta t} &\approx \prod_{j} e^{-\mathrm{i} h_j\delta t} \nonumber \\
&= \prod_{j} V_j^\dagger\, e^{-\mathrm{i} D_j \delta t}  V_j.
\end{align}
Finally, we can employ the ``tree algorithm'' of 
Ref.~\cite{Murairi:2022zdg} to implement an exact circuit for $e^{-\mathrm{i} D_j \delta t}$ using relatively few two-qubit gates. Namely, we demand only 236 {CNOT} gates here for fuzzy gauge theory, as opposed to $\sim$ 17000 {CNOT} gates entailed by simulating Horn's model of $SU(2)$ gauge theory~\cite{HORN1981149} for example.

\bibliographystyle{apsrev4-1}
\bibliography{paper-refs.bib}

\end{document}